\documentclass[11pt,a4paper]{article}

\usepackage{amsfonts}
\usepackage{graphicx}

\title{\textbf{Quantum magnetic flux lines, BPS vortex \\ zero modes,  and one-loop string tension shifts}}

\author{A. Alonso Izquierdo$^{(a)}$, J. Mateos Guilarte$^{(b)}$, and M. de la Torre Mayado$^{(b)}$
\\ {\normalsize {\it $^{(a)}$ Departamento de Matematica
Aplicada}, {\it University of Salamanca, SPAIN}} \\ {\normalsize {\it $^{(a)}$ Departamento de Fisica
Fundamental}, {\it University of Salamanca, SPAIN}}}

\date{}

\begin{document}

\maketitle

\begin{abstract}
Spectral heat kernel/zeta function regularization procedures are employed in this paper to control the divergences arising from
vacuum fluctuations of Bogomolnyi-Prasad-Sommerfield vortices in the Abelian Higgs model. Zero modes of vortex fluctuations are
the source of difficulties appearing when the standard Gilkey-de Witt expansion is performed. A modified GdW expansion is developed
to diminish the impact of the infrared divergences due to the vortex zero modes. With this new technique at our disposal we compute
the one-loop vortex mass shift in the planar AHM and the quantum corrections to the string tension of the magnetic flux tubes living in three dimensions. In both cases it is observed that weak repulsive forces surge between these classically non interacting topological defects caused by vacuum quantum fluctuations.
\end{abstract}

PACS: 11.15.Kc; 11.27.+d; 11.10.Gh

\section{Introduction}

Magnetic flux tubes with vortex filaments at their core were discovered by Abrikosov in the Ginzburg-Landau theory of Type II superconductivity \cite{Abrikosov1957:spj}. In that context these extended string like objects are macroscopic and do not require an specific treatment in a quantum framework. Nielsen and Olesen, however, rediscovered identical extended objects in the relativistic Abelian Higgs model, see \cite{Nielsen1973:npb}, and proposed
for them to play a r$\hat{\rm o}$le in hadronic physics as dual strings. It is thus clear after the Nielsen-Olesen proposal that in this new framework the vortex filaments are of quantum nature and there is the need of
clarifying to what kind of quantum state they correspond. It was later shown by Bogomolnyi \cite{Bogomolny1976:sjnp,Prasad1975:prl} that Abrikosov-Nielsen-Olesen vortices, seen in a two dimensional space, belong to a special class of topological solitons when the masses of the scalar and vector particles in the AHM are equal, or, the correlation lengths of scalar and magnetic fields correspond to the critical point between Type I and Type II phases
in Ginzburg-Landau superconductors.

It is thus natural to try the understanding of quantum Bogomolnyi-Prasad-Sommerfield planar vortices in the framework of the general quantum theory of solitons. The first successful attempts in this direction were achieved by Vassilevich in \cite{Vassilevich2003:prd}, and Rebhan et al. in \cite{Rebhan2004:npb}, by attacking this problem in the $N=2$ supersymmetric AHM. Almost simultaneously Bordag and Drozdov in \cite{Bordag2003:prd}
computed the vacuum energy due to purely fermionic fluctuations on a Nielsen-Olesen vortex. Together with other colleague we performed similar calculations in the purely bosonic planar AHM in References \cite{Alonso2004:prd} and \cite{Alonso2005:prd}. We used the spectral heat kernel/zeta function regularization procedure to control the divergences, both ultraviolet and infrared, arising in the computation of vacuum energies caused by one-loop fluctuations of BPS vortices, as well as those associated to tadpole and self-energy graphs. Invented by Hawking \cite{Hawking1977:cmp} and Dowker et al \cite{Dowker1976:prd} to describe quantum  fields in curved space-times this method was used for the first time in the analysis of quantum fluctuations of kinks and solitons by van Nieuwenhuizen et al in \cite{Bordag2002:prd} within a $N=2$ supersymmetric framework. We took profit of these ideas to calculate the quantum corrections to the masses of several types of topological kinks in scalar field models with different number of fields in References \cite{Alonso2002:npb,Alonso2002:npb2,Alonso2004:npb}.

The vortex Casimir energy is the main ingredient in the formula giving the vortex mass (2D) or string tension (3D) quantum corrections.
It is formally given by the trace, both in the matrix and the $L^2$-functional sense, of the square root of the matrix elliptic partial differential operator that governs the one-loop vortex fluctuations. This Hessian operator is a matrix second-order partial differential operator (PDO) of Schr$\ddot{\rm o}$dinger type. Its square root is defined in the framework of complex powers of elliptic (pseudo) differential operators, a well developed and sound mathematical theory. The formal trace is then the spectral zeta function of the elliptic PDO exhibiting analytical properties in the complex $s$-plane of the exponent. Nevertheless, use of the zeta function with the purpose of regularizing divergences in QFT requires to dispose of more detailed information about its description. The usual strategy developed by the physicist's community is to take profit of the more tractable spectral heat function to pass to the zeta function via Mellin transform, see e.g. References \cite{Elizalde1994,Vassilevich2003:prc,Avramidi2002:npp}. In particular it is a common technique in dealing with quantum fields on curved spaces and/or extended/solitonic backgrounds to start from the high-temperature (short time) asymptotic expansion of the heat equation kernel following the seminal works of deWitt \cite{deWitt1965} and Gilkey \cite{Gilkey1975:jdg} {\footnote{A lucid discussion of the differences between deWitt and Gilkey approaches may be found in the textbook \cite{Vassilevich2011} by Fursaev and Vassilevich.}}. All this machinery is well behaved if the field fluctuations are strictly $L^2$. In QFT, however, two characteristics of the spectrum of the PDO at the stake disturb this naif picture:
(1) First, usually there are fluctuations belonging to the continuous spectrum. To cope with this problem one put the system in a normalization box and impose periodic boundary conditions on the fields. Equivalently, a toric variety is taken as space and only at the end the volume is allowed to go to infinity. (2) Second, much more dangerous is the existence of massless particles and/or zero mode fluctuations. These long range fluctuations do not disappear in the low temperature (long time) regime and use of the high temperature asymptotics is made dubious. Barvinsky and Vilkovisky proposed to introduce non-local terms to treat this problem in covariant perturbation theory, see e.g. Reference \cite{Barvinsky1990:npb}, an idea that was put at work by Gusev and Zelnikov \cite{Gusev2000:prd} to compute the effective action in dilatonic two-dimensional gravity. Recall that effective actions are related to determinants of elliptic PDO, susceptible of being regularized by means of the derivative of the spectral zeta function at the origin of the $s$-complex plane.

E. Weinberg in \cite{Weinberg1979:prd} showed an index theorem in the open $\mathbb{R}^2$-plane for the deformation operator arising from the linear perturbations of the first-order partial differential equations
satisfied by self-dual/BPS vortices. The theorem, see also \cite{Weinberg2012}, stated that the algebraic kernel of the deformation operator has dimension $2 N$ where $N$ is the number of quanta of magnetic flux (vorticity) carried by the vortex solution. This means that there exist $2 N$
zero modes of fluctuation around BPS vortices linearly independent. Our main goal in this paper is to compute the quantum correction to the BPS vortex string tension induced by the vortex fluctuations taking into account
the existence of these vortex zero modes. Essentially we shall follow an strategy similar to that developed in \cite{Barvinsky1990:npb} and \cite{Gusev2000:prd} but we shall adapt our treatment to the heat kernel/zeta function procedure
as applied in quantum theory of solitons. Specifically, our new technique is tailored in order to incorporate the impact of zero modes in the infrared in the Gilkey-deWitt heat kernel expansion. In fact, in Reference \cite{Alonso2012:epjc} we proposed and tested the improved heat kernel expansion, with the impact of zero modes under control, in scalar one-field theory models in order to
compute one-loop kink mass shifts. Limitations in the use of the standard GdW procedure arise when zero modes enter the game because the asymptotic low temperature behaviour of the heat function
cannot be reproduced and we were forced to restrict the Mellin transform to a finite range near the high temperature regime. The contributions of the low energy fluctuations to the spectral zeta function are thus almost suppressed. In this sense the question about if the quantum fluctuations induce forces between the BPS vortices remained unsolved because of the lack of control on the previously mentioned source of errors.

The idea to repair this difficulty was to include in the heat kernel expansion a (non local) term that takes care of the effect of zero modes surviving in the low temperature range. The new term induced by the zero modes depends of an arbitrary a priori function of the (fictitious) temperature which is chosen by demanding two properties: (1) The known behaviour of the heat kernel not only at high but also at low temperature are reproduced. (2) The solution of the recurrence relations implied by the asymptotic expansion is minimally perturbed by the arbitrary function. This modification allowed a much more precise evaluation of the Mellin's transform of the heat trace to obtain the spectral zeta extending the integration interval to all the temperature range. By this token we are able to fix not only the zeta function near the poles but also the entire part. Because in the kink case many exact evaluations of kink mass quantum corrections are known we were able to check that the improved heat kernel expansion offered much closer approximations to the exact results as compared with the results obtained by using the standard GdW method. Moreover, in Reference \cite{Alonso2014:jhep} we extended the procedure to many
component scalar field theory. In these type of models there are families of BPS kinks in such a way than other kink zero modes besides the translational zero mode arises. The results also were much more precise than those previously obtained using the standard GdW expansion, see e.g. \cite{Alonso2006:hepth} and \cite{Mateos2009:pos}. More interesting, in this last paper we do not only consider the problem in $(1+1)$-dimensional space-time but we analyzed the one-loop fluctuations in a three dimensional perspective where kinks become domain walls. In Reference \cite{Rebhan2002:njp} the same problem was addressed over standard supersymmetric kink domain walls relying on dimensional regularization procedures.
Our method consequently also works for extended objects of $p$-brane type and, in the case of the model we studied, an interesting phenomenon was unveiled: within a family of classically degenerate BPS kinks repulsive forces were induced by the quantum fluctuations that broke the classical degeneracy. We plan to address an identical issue in the moduli space of BPS vortices in the Abelian Higgs model. Jaffe and Taubes showed in \cite{Jaffe1980} that the vortex moduli space is the set of $N$ unordered points in $\mathbb{R}^2$.
As a consequence vortices with one quantum of magnetic flux move freely without any interaction. The $2N$ vortex zero modes obey to this freedom in the critical point between Type II and Type I superconductivity phases, in the first case surge repulsive whereas in the second case attractive forces between vortices. We shall discuss wether or not this classical picture is maintained at one-loop order after the effect of vortex fluctuations is accounted for. We shall perform the pertinent calculations generalizing the improved GdW heat kernel procedure developed previously for scalar field theories to Abelian gauge theories with spontaneous symmetry breaking. The analysis will be first focused in the planar AHM where the BPS vortices are topological solitons. After that we shall move to study the same problem in a three dimensional space, where we find BPS vortices magnetic tubes or vortex strings. To perform this task, evaluation of the quantum corrections to BPS vortices by using the modified GdW expansion we need a detailed information of the spectrum of the matrix second-order PDO
that governs the vortex fluctuations. All the information needed about vortex zero modes and bound states is collected in our recent papers \cite{Alonso2016:plb} and \cite{Alonso2016:jhep} where pertinent References can be found.

The organization of the paper is as follows: In Section \S.2 we thoroughly address the problem described above in the planar system.  Subsection \S.2.1 summarize the well known facts about planar BPS fluctuations, in sub-Section \S.2.2 the modified or improved GdW heat kernel expansion is generalized to planar
Abelian gauge systems with spontaneous symmetry breaking by one scalar field, and finally, in sub-Section \S.2.3 the one-loop mass shifts of rotationally symmetric planar BPS vortices are computed. The new one-loop mass vortex shifts performed in this work, although qualitatively compatible with those obtained in \cite{Alonso2005:prd}, are of much greater precision because the new technique is able to incorporate also the effect of zero modes in the spectral zeta function.  Section \S.3 is fully devoted to describe the quantum corrections at one loop order of the BPS vortex string tensions in the three dimensional Abelian Higgs model. These results are completely new. Some conclusions about the new approach developed here and the induction of repulsive forces between BPS due to their quantum fluctuations offered in Section \S.4, where further comments on possible generalizations/extensions of this problem to other physical scenarios are elaborated.

\section{One-loop mass shifts of rotationally symmetric planar BPS vortices}

\subsection{Quantum fluctuations of BPS vortices in the planar Abelian Higgs model}

The Abelian Higgs model describes the minimal coupling between an $U(1)$-gauge field and a complex scalar field in a phase where the gauge symmetry is spontaneously broken. In terms of non-dimensional variables, $x^\mu \rightarrow \frac{1}{e v}x^\mu$, and fields, $\phi\rightarrow v\phi$, $A_\mu \rightarrow v A_\mu$, where $e$ and $v$ are respectively the gauge coupling and the modulus of the vacuum expectation value of the scalar field, the action functional for the AHM in (2+1)-dimensions reads
\begin{equation}
S[\phi,A]=\frac{v}{e}\int d^3x \left[ -\frac{1}{4} F_{\mu\nu}F^{\mu \nu} + \frac{1}{2} (D_\mu \phi)^* D^\mu \phi -\frac{\kappa^2}{8} (\phi^* \phi-1)^2 \right] \, \, \, .
\label{action1}
\end{equation}
The main ingredients entering this formula are: the complex scalar field $\phi=\phi_1+i\phi_2$, the vector gauge potential $A=(A_0,A_1,A_2)$, the covariant derivative $D_\mu \phi = (\partial_\mu -i A_\mu)\phi$ and the field tensor $F_{\mu\nu}=\partial_\mu A_\nu - \partial_\nu A_\mu$. We choose a systems of units where $c=1$, but $\hbar$ has dimensions of length$\times$mass. The metric tensor in the Minkowski space $\mathbb{R}^{(2,1)}$ is chosen as $g_{\mu\nu}={\rm diag}(1,-1,-1)$ with $\mu,\nu=0,1,2$. The parameter $\kappa^2=\frac{\lambda}{e^2}$, where $\lambda$ is the quartic self-coupling of the scalar field, measures the ratio between the square of the masses of the Higgs, $M^2=\lambda v^2$, and the vector particles, $m^2=e^2 v^2$. Bogomolnyi-Prasad-Sommerfield (self-dual) vortices arise when the parameter $\kappa^2$ is set to unity, $\kappa^2=1, $ in the action (\ref{action1}). These vortices are solitonic topological defects (static and spatially localized solutions of the field equations) for which the static energy density functional
\[
V[\phi,A]=v^2\int d^2 x \Big[ \frac{1}{4} F_{ij} F_{ij} + \frac{1}{2} (D_i\phi)^* D_i\phi + \frac{1}{8} (\phi^*\phi-1)^2 \Big]
\]
is finite. A Bogomolnyi arrangement of $V[\phi,A]$
\begin{equation}
V[\phi,A]=\frac{v^2}{2}\int_{\mathbb{R}^2} \, d^2x \, \left\{\Big(F_{12}\pm\frac{1}{2}(\phi^*\phi-1)\Big)^2+\big\vert D_1\phi\pm i D_2 \phi\vert^2\right\}+\frac{v^2}{2} \,\,\Big\vert \int_{\mathbb{R}^2} \, d^2x \, F_{12}\Big\vert \label{bogs}
\end{equation}
leads us to conclude that solutions of the first-order PDE system
\begin{equation}
D_1\phi \pm i D_2 \phi =0 \hspace{0.5cm},\hspace{0.5cm} F_{12}\pm \frac{1}{2} (\phi^*\phi -1)=0 \label{pde1}
\end{equation}
complying with the asymptotic boundary conditions
\begin{equation}
\phi^* \phi|_{S_\infty^1}=1 \hspace{0.5cm} \mbox{and} \hspace{0.5cm} D_i\phi|_{S_\infty^1}=0 \hspace{0.5cm} \equiv \hspace{0.5cm}
 \phi |_{S^1_\infty}=e^{i N \theta} \hspace{0.5cm} \mbox{and} \hspace{0.5cm} A_i|_{S^1_\infty}=-i N\phi^*\partial_i \phi|_{S^1_\infty} \label{asymptotic}
\end{equation}
where $\theta =\arctan \frac{x^2}{x^1}$, $S_r^1= \{(x^1,x^2):  x^1 x^1+x^2 x^2=r^2 \}$ and $S_\infty^1=\lim_{r\to +\infty} S_r$, have a classically quantized magnetic flux: $\frac{1}{2\pi}\int_{\mathbb{R}^2}\, dx^1dx^2 \, F_{12}(x^1,x^2)=N\in\mathbb{Z}$.

It is clear from (\ref{asymptotic}) that the vector field $A_i$ is asymptotically purely vorticial. Jaffe and Taubes, showed, see \cite{Jaffe1980}, that the solutions are determined from $N$ points freely located in the $\mathbb{R}^2$ plane, around each of which the vector field $A_i$ is a quantized vortex, the total magnetic charge being equal to $N$. It is well known that these magnetically charged objects are also solutions of the second-order static field equations and, because they satisfy the PDE system (\ref{pde1}), the BPS vortices are absolute minima of the action in the different topological sectors characterized by $N$. Therefore, these BPS or self-dual vortices are stable. The upper signs in (\ref{pde1}) refer to the topological defects with a positive winding number, $N>0$, (vortices), of the map from the $\mathbb{R}^2$ boundary circle $S_\infty^1$ at infinity to the vacuum circle $S^1_1$ determined by the asymptotic behaviour of the complex field, see (\ref{asymptotic}). Solutions of (\ref{pde1}) with the lower signs are topological defects with negative winding number, $N<0$, (anti-vortices).

Without loss of generality, we shall focus in this paper on solutions with positive magnetic charge, although an identical analysis could be easily developed for anti-vortices. We shall denote by
\[
\psi(\vec{x};N)=\psi_1(\vec{x};N) + i \, \psi_2(\vec{x};N) \hspace{0.5cm},\hspace{0.5cm} V(\vec{x};N)=(V_1(\vec{x};N),V_2(\vec{x};N))
\]
the scalar and vector fields of the BPS vortex solutions; the vorticity number, the magnetic charge, will be specified if necessary. Perturbations of these vortex classical solutions in the form
\begin{equation}
\widetilde{A}_i(\vec{x};N)= V_i(\vec{x};N) +\epsilon \, a_i(\vec{x}) \hspace{0.5cm},\hspace{0.5cm} \widetilde{\phi}_i(\vec{x};N)=\psi_i(\vec{x};N) + \epsilon\, \varphi_i(\vec{x}) \hspace{0.5cm},\hspace{0.5cm} i=1,2 \label{perturbed}
\end{equation}
respond to small fluctuations around the topological defects and open a window to observe the behaviour of these objects in the quantum world up to the semi-classical or one-loop order.

The analysis of the physics of the BPS vortex small fluctuations starts by assembling them in a four-component column which we write as the transpose of the four-component field vector
\[
\xi(\vec{x})=\left( \begin{array}{cccc}a_1(\vec{x}) & a_2(\vec{x}) & \varphi_1(\vec{x}) & \varphi_2(\vec{x}) \end{array} \right)^{\rm t}\, \, .
\]
In order to avoid spurious pure gauge fluctuations we impose the background gauge condition
\begin{equation}
B(a_k,\varphi,\phi)=\sum_{k=1}^2 \partial_k a_k-(\psi_1 \varphi_2-\psi_2\varphi_1)=0
\label{backgroundgauge}
\end{equation}
which can be generated as a field equation by adding to the action the following gauge fixing term:
\[
S^{(GF)}=\frac{1}{2} \int d^3 x [B(a_k,\varphi,\psi)]^2 \, \, .
\]
The expansion of the action up to quadratic order in the fluctuations plus the gauge fixing term reads
\[
\delta^{(2)}S+S^{(GF)}= -\frac{v}{e}\int_{\mathcal{R}^{2,1}} \, dx^0 dx^1 dx^2\,\left\{ \xi^{\rm t}(x^0,\vec{x})\left[\frac{\partial^2}{\partial (x^0)^2}+{\cal H}^+\right] \xi(x^0,\vec{x})\right\}+o(\xi^3)
\]
where
\begin{equation}
{\cal H}^+= \left( \begin{array}{cccc}
-\Delta + |\psi|^2 & 0 & -2D_1 \psi_2 & 2 D_1 \psi_1 \\
0 & -\Delta +|\psi|^2 & -2 D_2 \psi_2 & 2 D_2 \psi_1 \\
-2 D_1 \psi_2 & -2 D_2\psi_2 & -\Delta +\frac{1}{2} (3|\psi|^2-1)+V_kV_k & -2 V_k \partial_k -\partial_k V_k \\
2D_1\psi_1 & 2 D_2 \psi_1 & 2V_k \partial_k + \partial_k V_k & -\Delta +\frac{1}{2} (3|\psi|^2-1) + V_kV_k
 \end{array} \right) \label{operator1}
\end{equation}
is the Hessian or second-order fluctuation operator and terms of third and quartic order in the perturbations are neglected. In the operator ${\cal H}^+$ we denote: $D_1\psi_1 = \partial_1 \psi_1 + V_1 \psi_2$, $D_2\psi_1 = \partial_2 \psi_1 + V_2 \psi_2$, $D_1\psi_2 = \partial_1 \psi_2 - V_1 \psi_1$ and $D_2\psi_2 = \partial_2 \psi_2 - V_2 \psi_1$.
In this background gauge the classical energy up to the quadratic order in the small fluctuations is now easily derived:
\[
H^{(2)}+H^{(GF)}=\frac{v^2}{2}\int_{\mathbb{R}^2}\, dx^1dx^2 \, \left\{\frac{\partial\xi^t}{\partial t}(\vec{x},t)\frac{\partial\xi}{\partial t}(\vec{x},t)+\xi^t(\vec{x},t){\cal H}^+\xi(\vec{x},t)\right\} +o(\xi^3)\, .
\]
We impose finiteness of the norm on the static fluctuations, equivalently the fixed time perturbations, $\xi(\vec{x})$: $\|\xi(\vec{x})\|^2  = \int_{\mathbb{R}^2} d^2x [ (a_1(\vec{x}))^2 + (a_2(\vec{x}))^2 + (\varphi_1(\vec{x}))^2 + (\varphi_2(\vec{x}))^2 ] < +\infty$. Thus, the four component vectors of real functions $\xi(\vec{x})$ belong to the Hilbert space of square integrable vector functions, $\xi(\vec{x})\in \oplus_{a=1}^{4}\, L^2_a(\mathbb{R}^2)$. A previous step to quantize this system is to perform
the \lq\lq normal mode\rq\rq{} expansion, i.e., use the eigenvectors of ${\cal H}^+$ as a base to expand the fluctuations:
\begin{equation}
{\cal H}^+ \xi_\omega(\vec{x})=\omega^2 \xi_\omega(\vec{x}) \, \, , \, \, \omega^2\geq 0 \quad , \qquad \quad \xi(\vec{x},t)= \int[d\omega] \, e^{i \omega t} a^t(\omega) \, \xi_\omega(\vec{x}) \label{fluctspec} \, \, .
\end{equation}
It is well known, see \cite{Alonso2016:jhep} and References quoted therein to find a summary, that the ${\cal H}^+$ operator has a kernel of dimension $2 N$, i.e., there are $2 N$ lineally independent eigenfunctions of zero eigenvalue in the spectrum of ${\cal H}^+$. There is also a discrete set of eigenfunctions with positive eigenvalues but lesser than one: $0<\omega^2<1$. These are eigenfunctions of ${\cal H}^+$ where the positive fluctuations are trapped in bound states at the vortex core. Finally, there are eigenfunctions in the continuous spectrum of ${\cal H}^+$ with threshold precisely at $\omega^2=1$. In formula (\ref{fluctspec}) the $a^t(\omega)$-coefficients describe the four-vector normal modes of fluctuation and the integration symbol $\int [d\omega]$ means that the expansion encompasses both fluctuations in the pure point spectrum and those in the continuous spectrum.

It is interesting at this point to summarize a heat function proof of the Atiyah-Singer-Weinberg index theorem \cite{Weinberg1979:prd,Weinberg2012}. Weinberg showed the existence of $2N$ linearly independent zero modes $\xi_0(\vec{x})$ of ${\cal H}^+$ (eigenfunctions with zero eigenvalues). Weinberg's proof rely on a supersymmetric structure built on perturbations of solutions of  (\ref{pde1}) which are still solutions. Perturbing the PDE (\ref{pde1}) system of three equations together with the background gauge
one finds that new solutions arise, complying with the background gauge, if and only if the perturbations belong to the kernel of the
deformation operator ${\cal D}$:
\begin{equation}
\hspace{-2cm}{\cal D}= \left( \begin{array}{cccc}
-\partial_2 & \partial_1 & \psi_1 & \psi_2 \\
-\partial_1 & -\partial_2 & -\psi_2 & \psi_1 \\
\psi_1 & -\psi_2 & -\partial_2 + V_1 & -\partial_1 -V_2 \\
\psi_2 & \psi_1 & \partial+V_2 & -\partial_2 + V_1
\end{array} \right) \quad ,\quad \quad \hspace{1cm}{\cal D}\xi_0(\vec{x})=0 \label{zeromode1} \, \, .
\end{equation}
It is easy to check that the Hessian ${\cal H}^+$ factorizes as the product of the ${\cal D}$ operator times its adjoint:  ${\cal H}^+={\cal D}^\dagger \, {\cal D}$. Besides of showing that the four-vector columns $\xi_0(\vec{x})$ are zero modes of ${\cal H}^+$, this factorization hides a supersymmetric quantum mechanical structure where the partner Hamiltonian of ${\cal H}^+$ is:
\[
{\cal H}^- = {\cal D} \, {\cal D}^\dagger= \left( \begin{array}{cccc}
-\Delta + |\psi|^2 & 0 & 0 & 0 \\
0 & -\Delta +|\psi|^2 & 0 & 0 \\
0 & 0 & -\Delta +\frac{1}{2} (|\psi|^2+1)+V_kV_k & -2 V_k \partial_k -\partial_k V_k \\
0 & 0 & 2V_k \partial_k + \partial_k V_k & -\Delta +\frac{1}{2} (|\psi|^2+1) + V_kV_k
 \end{array} \right)
\]
${\cal H}^+$ and ${\cal H}^-$ are isospectral operators (although the spectral densities in the continuous spectra may differ). Thus, the index of ${\cal D}$, regularized by means of the spectral heat functions of ${\cal H}^\pm$,
\[
{\rm ind}\,{\cal D}= {\rm dim}\,{\rm Ker}\,{\cal D}-{\rm dim}\,{\rm Ker}\,{\cal D}^\dagger={\rm Tr}_{L^2}\, e^{-\beta {\cal H}^+} - {\rm Tr}_{L^2}\, e^{-\beta {\cal H}^-} \ \, ,
\]
where $\beta$ is a fictitious inverse temperature, is independent of $\beta$. It is possible to evaluate the difference between the functional traces in the $\beta=0$ limit having in mind that
the operators ${\cal H}^\pm$ have the structure of Schr$\ddot{\rm o}$dinger operators: ${\cal H}^\pm = {\cal H}_0+ \vec{\mathbf{Q}}(\vec{x})\cdot \vec{\nabla} + \mathbf{U}^\pm(\vec{x})$, where ${\cal H}_0$ is the Helmoltz operator times the $4\times 4$ unit matrix and the matrix potentials read:
{\small\begin{equation}
\mathbf{U}^\pm(\vec{x})= \left(\begin{array}{cccc}\vert\psi\vert^2-1 & 0 & -(D_1\psi_2\pm D_1 \psi_2) &  D_1\psi_1\pm D_1 \psi_1\\ 0 & \vert\psi\vert^2-1 & -(D_2\psi_2\pm D_2 \psi_2) & D_2\psi_1\pm D_2 \psi_1\\ -(D_1\psi_2\pm D_1 \psi_2) & -(D_2\psi_2\pm D_2\psi_2) & (1\pm\frac{1}{2})(\vert\psi\vert^2-1)+V_kV_k & 0 \\ D_1\psi_1\pm D_1 \psi_1 & D_2\psi_1\pm D_2 \psi_1 & 0 & (1\pm\frac{1}{2})(\vert\psi\vert^2-1)+V_kV_k
\end{array}\right). \label{vspot}
\end{equation}}
Use of the high-temperature heat trace asymptotic expansions,
\[
{\rm Tr}_{L^2}{\rm exp}(-\beta{\cal H}^\pm)\simeq \frac{e^{-\beta}}{4\pi} \sum_{n=1}^\infty \, {\rm tr} [ c_n({\cal H}^\pm)]\beta^{n-1}
\]
leads to estimate the index in the form:
\[
{\rm ind}\,{\cal D}=\lim_{\beta\to 0} \left({\rm Tr}_{L^2}\, e^{-\beta {\cal H}^+} - {\rm Tr}_{L^2}\, e^{-\beta {\cal H}^-}  \right)=\frac{1}{4\pi}\left({\rm tr} [ c_1({\cal H}^+)]-{\rm tr}[ c_1({\cal H}^-)] \right).
\]
Here ${\rm tr}$ refers to the conventional $(4\times 4)$-matrix trace and the divergent ${\rm tr} [c_0({\cal H}^\pm)]$ terms have been discarded because they cancel each other in the index formula. We shall see that
\begin{equation}
{\rm tr}[ c_1({\cal H}^\pm)]=-{\rm tr} \int_{\mathbb{R}^2} \, d^2x \, U^\pm(\vec{x}) \, \, , \quad  \, \, \mbox{henceforth}, \, \,  \, \, \quad
{\rm ind}\,{\cal D}=\frac{1}{2 \pi} \int_{\mathbb{R}^2} \, d^2x \, \left(1-\vert \psi\vert^2\right)= 2 N \label{vind} \, \, .
\end{equation}
But ${\cal H}^-$ is a positive definite operator such that ${\rm dim}\,{\rm Ker}\,{\cal D}^\dagger=0$, which means that ${\cal H}^+$ has $2N$ zero modes.

\subsection{BPS vortex heat kernel asymptotic expansion: impact of zero modes}

One of the main goals in this paper is to compute one-loop vortex mass shifts. In References \cite{Alonso2004:prd,Alonso2005:prd,Alonso2008:npb}, see also the reviews \cite{Alonso2006:hepth,Mateos2009:pos}, we performed these calculations by applying the spectral zeta function regularization procedure to the second-order small vortex fluctuation operator ${\cal H}^+$ both in the Abelian Higgs model and in Semilocal Abelian gauge systems. The scheme developed by our group was based in the standard Gilkey-de Witt heat kernel asymptotic expansion. An important obstacle found in developing this program is that the Gilkey-de Witt approach is well established only for operators with strictly positive spectrum and the operator ${\cal H}^+$ exhibits zero modes. In the papers \cite{Alonso2012:epjc,Alonso2014:jhep} two of us improved on the Gilkey-de Witt expansion by showing how to generalize the method to cope with the existence of zero modes. Application of the generalized Gilkey-de Witt heat kernel asymptotic expansion to the computation of one-loop kink mass shifts showed a remarkably better precision and unveiled the appearance of forces between kinks of pure quantum nature.

In this Section, having in the back of the mind computations of BPS vortex mass shifts, we shall generalize the standard Gilkey-de Witt heat kernel expansion to operators with zero modes in its spectrum within the class of the BPS vortex Hessian operator ${\cal H}^+$. The new development is one of the main novel proposals in this paper. With this objective in mind, but looking at a larger class of operators containing ${\cal H}^+$, we consider a general second-order $D\times D$ matrix PDO of the form
\begin{equation}
{\cal H} = - \textbf{I} \, \Delta + \textbf{u}^2 + \textbf{U}(\vec{x}) + \vec{\textbf{Q}} (\vec{x}) \cdot \vec{\nabla} \label{generaloperator}
\end{equation}
where $\textbf{I}$ is the $D\times D$ identity matrix, $\Delta= \frac{\partial^2}{\partial x_1^2} + \frac{\partial^2}{\partial x_2^2}$ is the 2D Laplacian, $\textbf{u}= {\rm diag}\,\{u_1,\dots,u_D\}$ is a constant $D\times D$ diagonal matrix determined by the asymptotic behaviour, $\vert \vec{x}\vert\to \infty$, of ${\cal H}$ and $\textbf{U}(\vec{x}) = (U_{ab}(\vec{x}))_{D\times D}$ with $a,b=1,2, \dots, D$, is a $D\times D$-matrix potential well. Besides $\vec{\textbf{Q}} (\vec{x}) =(\textbf{Q}_1(\vec{x}),\textbf{Q}_2(\vec{x}))$ is a vector field of matrices such that the last term in (\ref{generaloperator}) reads
\[
\vec{\textbf{Q}} (\vec{x}) \cdot \vec{\nabla} = \Big(\sum_{i=1}^2 [\textbf{Q}_i(\vec{x})]_{ab} \, \partial_i\Big)_{D\times D}
\]
We assume that
\begin{equation}
\lim_{\vert\vec{x}\vert\rightarrow +\infty} \textbf{U}(\vec{x}) = \textbf{0} \hspace{0.5cm},\hspace{0.5cm} \lim_{\vert\vec{x}\vert\rightarrow +\infty} \textbf{Q}(\vec{x}) = \textbf{0} \label{generaloperator0}
\end{equation}
which implies that the operator (\ref{generaloperator}) asymptotically behaves as the PDO ${\cal H}_0=- \textbf{I} \, \Delta+ \textbf{u}^2 $. It is direct to check that the second-order small vortex fluctuation operator is encompassed in formula (\ref{generaloperator}) for $D=4$ and the following assignments of vacuum diagonal matrix and first-order PDO vector field
\[
\textbf{v}={\rm diag}\,\{1,1,1,1\} \quad , \quad \quad
\textbf{Q}_k(\vec{x})= \left( \begin{array}{cccc} 0 & 0 & 0 & 0 \\ 0 & 0 & 0 & 0 \\ 0 & 0 & 0 & -2V_k \\ 0 & 0 & 2V_k & 0 \end{array} \right)\, \, ,
\]
whereas the $4\times 4$-matrix potential well is defined in (\ref{vspot}).

The Gilkey-de Witt approach aims to construct a power series expansion of the ${\cal H}$-spectral heat trace $h_{\cal H}(\beta) = {\rm Tr}_{L^2} \, e^{-\beta \, {\cal H}}$
by taking advantage of the fact that this function can be obtained from integration all over the plane of the diagonal ${\cal H}$-heat equation kernel:
\begin{equation}
h_{{\cal H}}(\beta) = \int_{\mathbb{R}^2}\, d^2 x \,\, {\rm tr}\, \textbf{K}_{\cal H} (\vec{x},\vec{x};\beta) \label{heatfunction0}
\end{equation}
i.e., the trace in both the $L^2$-functional and $D\times D$-matrix senses of the integral kernel of the ${\cal H}$-heat equation:
\begin{equation}
\Big( \frac{\partial}{\partial \beta} + {\cal H} \Big) \mathbf{K}_{\cal H} (\vec{x},\vec{y},\beta) =0 \hspace{0.5cm} , \hspace{0.5cm} \mathbf{K}_{\cal H}(\vec{x},\vec{y};0)=\delta^{(2)}(\vec{x}-\vec{y} ) \, \textbf{I}_{D\times D} \, \, . \label{heatequation}
\end{equation}
Completeness of the eigenfunctions of ${\cal H}$ allows to write the fundamental solution of equation (\ref{heatequation}) as the expansion
\begin{equation}
\textbf{K}_{\cal H}(\vec{x},\vec{y};\beta)= \sum_{\ell=1}^{N_{\rm zm}} \Xi_{0\ell}(\vec{x}) \, \Xi_{0\ell}^\dagger (\vec{y}) + \sum_{n=1}^{N_B} \Xi_n(\vec{x}) \, \Xi_n^\dagger (\vec{y}) e^{-\beta \omega_n^2} + \int [dk_1 dk_2] \, \Xi_{\vec{k}}(\vec{x}) \, \Xi_{\vec{k}}^\dagger(\vec{y}) \, e^{-\beta \omega^2(\vert\vec{k}\vert)} \label{integralkernel01}
\end{equation}
Here $N_{\rm zm}$ denotes the number of zero modes $\Xi_{0\ell}(\vec{x})$,  $N_{\rm zm}$ linearly independent functions belonging to the algebraic kernel of ${\cal H}$, $N_B$ is the number of bound states $\Xi_n(\vec{x})$ in ${\rm Spec}({\cal H})$, and $\Xi_{\vec{k}}(\vec{x})$ are the continuous spectrum eigenfunctions of the operator ${\cal H}$. They are $D$-component functions and form an orthonormal basis in the Hilbert space $\oplus_{a=1}^D L^2_a(\mathbb{R}^2)$. The $\beta=0$ (infinite temperature) condition in (\ref{heatequation}) is derived from the completeness of the set of ${\cal H}$-eigenfunctions.

The standard Gilkey-de Witt cunning strategy is based in using the knowledge of the ${\cal H}_0$ heat kernel. In a normalizing square of
area $L^2$ it reads:
\begin{equation}
\mathbf{K}_{{\cal H}_0} (\vec{x},\vec{y};\beta)= \frac{l^2}{4\pi \beta} \, e^{-\frac{\|\vec{x}-\vec{y}\|^2}{4\beta}} \, e^{-\beta \textbf{u}^2} \hspace{0.5cm}, \hspace{0.5cm} e^{-\beta \textbf{u}^2}= {\rm diag}\,\{e^{-\beta u_1^2}, \dots, e^{-\beta u_D^2}\} \, \, \, , \, \, \, l=mL\label{hk0}
\end{equation}
and therefore in this context the ${\cal H}$ heat kernel is assumed to follow the factorization:
\begin{equation}
\textbf{K}_{\cal H}(\vec{x},\vec{y};\beta)= \textbf{A}(\vec{x},\vec{y};\beta) \, \textbf{K}_{{\cal H}_0} ( \vec{x},\vec{y};\beta) \label{factorizacion0}
\end{equation}
Plugging this ansatz into the heat equation (\ref{heatequation}) another equation for $\textbf{A}(\vec{x},\vec{y};\beta)$ (usually called transfer equation) arises that is solved by expanding $\textbf{A}$ as a power series in $\beta$.

This procedure is well behaved if the spectrum of the operator ${\cal H}$ is strictly positive provided that the infinite temperature condition $\textbf{A}(\vec{x},\vec{y};0)=\textbf{I}_{D\times D}$ is fixed. However, if the operator exhibits zero modes the factorization (\ref{factorizacion0}) is inconsistent because the left and right members in (\ref{factorizacion0}) behave in different ways at zero temperature, see (\ref{integralkernel01}) and (\ref{hk0}):
\begin{equation}
\lim_{\beta \rightarrow +\infty} \textbf{K}_{{\cal H}} (\vec{x},\vec{y}; \beta) = \sum_{\ell=1}^{N_{zm}} \Xi_{0\ell}(\vec{x}) \, \Xi_{0\ell}^\dagger (\vec{y}) \quad , \qquad \quad \lim_{\beta\rightarrow +\infty} \mathbf{K}_{{\cal H}_0} (\vec{x},\vec{y};\beta)=0 \label{Kinfinity} \, \, ,
\end{equation}
due to the fact that $\textbf{A}(\vec{x},\vec{y};\beta)$ grows as a power of $\beta$ when $\beta\to +\infty$. In order to amend this discrepancy we replace the factorization (\ref{factorizacion0}) by the following one:
\begin{equation}
\textbf{K}_{{\cal H}}(\vec{x},\vec{y};\beta)= \textbf{C}(\vec{x},\vec{y};\beta) \, \mathbf{K}_{{\cal H}_0} ( \vec{x},\vec{y};\beta) + \sum_{\ell=1}^{N_{\rm zm}} e^{-\frac{\|\vec{x}-\vec{y}\|^2}{4\beta}} \, \Xi_{0\ell}(\vec{x}) \, \Xi_{0\ell}^\dagger (\vec{y}) \,\, \textbf{G}(\beta) \, \, .
\label{factorizacion1}
\end{equation}
Good agreement between the zero temperature regime when zero modes are present, together the usual conditions at infinity temperature not affected by zero modes, are guaranteed provided that the matrix function $\textbf{G}(\beta)$ and the matrix density $\textbf{C}(\vec{x},\vec{y};\beta)$ satisfy:
\begin{equation}
\lim_{\beta \rightarrow +\infty} \textbf{G}(\beta)= \textbf{I}_{D\times D} \quad , \quad \quad \lim_{\beta\to 0} \textbf{G}(\beta)=0 \qquad \quad , \qquad \quad \lim_{\beta\to 0}\textbf{C}(\vec{x},\vec{y};\beta)= \textbf{I}_{D\times D} \label{initialconditionC}
\end{equation}
The matrix density $\textbf{C}(\vec{x},\vec{y};\beta)$, like $\textbf{A}(\vec{x},\vec{y};\beta)$ in the standard GdW method, relates the positive part of ${\rm Spec}{\cal H}$ to ${\rm Spec}{\cal H}_0$ in the ${\cal H}$-heat kernel, whereas the second term in the right hand side of (\ref{factorizacion1}) encodes the contribution of zero modes.

The power series expansion
\begin{equation}
\textbf{C}(\vec{x},\vec{y};\beta)=\sum_{n=0}^\infty \textbf{c}_n(\vec{x},\vec{y}) \, \beta^n \label{expansion5}
\end{equation}
together with the factorization (\ref{factorizacion1}) is plugged into the heat equation (\ref{heatequation}) as in the standard GdW procedure. The PDE (\ref{heatequation}) is converted thereafter into the following relations between the coefficients of the modified GdW expansion and their derivatives:
\begin{eqnarray}
&& -\,\frac{1}{2\beta} \,(\vec{x}-\vec{y}) \cdot \vec{\textbf{Q}}(\vec{x}) \,\, \textbf{c}_0(\vec{x},\vec{y}) + \sum_{n=0}^\infty \Big[(n+1)\textbf{c}_{n+1}(\vec{x},\vec{y}) - \Delta \textbf{c}_n(\vec{x},\vec{y})+ (\vec{x}-\vec{y})\cdot \vec{\nabla} \textbf{c}_{n+1}(\vec{x},\vec{y}) + \nonumber \\ && + \, \textbf{U}(\vec{x})\, \textbf{c}_n(\vec{x},\vec{y}) + [\textbf{u}^2,\textbf{c}_n(\vec{x},\vec{y})] + \vec{\textbf{Q}}(\vec{x}) \cdot \vec{\nabla} \textbf{c}_n(\vec{x},\vec{y}) - \frac{1}{2} \,(\vec{x}-\vec{y})\cdot \vec{\textbf{Q}}(\vec{x}) \,\,\textbf{c}_{n+1}(\vec{x},\vec{y}) \Big] \beta^n + \nonumber \\ && +\, \sum_{\ell=1}^{N_{\rm zm}} 4 \pi \Big[ \Xi_{0\ell}(\vec{x}) \, \Xi_{0\ell}^\dagger (\vec{y}) \Big( \beta \frac{d\textbf{G}}{d\beta}(\beta) + \textbf{G}(\beta) \Big) + (\vec{x}-\vec{y})\cdot \vec{\nabla} \Xi_{0\ell}(\vec{x}) \,\, \Xi_{0\ell}^\dagger (\vec{y}) \, \textbf{G}(\beta) - \nonumber\\ && -\,\frac{1}{2}\, (\vec{x}-\vec{y})\cdot \vec{\textbf{Q}}(\vec{x}) \,\, \Xi_{0\ell}(\vec{x}) \,\, \Xi_{0\ell}^\dagger (\vec{y}) \, \textbf{G} (\beta) \Big] e^{\beta \textbf{u}^2} =\textbf{0} \, . \label{mgdwrr}
\end{eqnarray}
Taking into account that eventually we shall take the limit $\vec{y}\rightarrow \vec{x}$ we can neglect the contribution of the first term in this relation. Before of attempting to solve (\ref{mgdwrr}) there is the need of selecting $\textbf{G}(\beta)$. Restricted by the zero and infinite temperature behaviours (\ref{initialconditionC}) and looking for optimizing the structure of (\ref{mgdwrr}) we choose:
\begin{equation}
\textbf{G}(\beta)=1-e^{-\beta \textbf{u}^2} \, \, . \label{zmgfunction}
\end{equation}
Substituting this $\textbf{G}(\beta)$ function into (\ref{mgdwrr}), expanding the lower two rows in (\ref{mgdwrr}) as a power series in $\beta$ and equalizing terms of the same power of $\beta$, a recurrence relation for the matrix densities $\textbf{c}_n(\vec{x},\vec{y})$ arises. We obtain
\begin{equation}
\textbf{c}_1(\vec{x},\vec{y}) - \Delta \textbf{c}_0(\vec{x},\vec{y}) + (\vec{x}-\vec{y})\cdot \vec{\nabla} \textbf{c}_1(\vec{x},\vec{y}) + \textbf{U}(\vec{x}) \,\, \textbf{c}_0(\vec{x},\vec{y})=\textbf{0} \label{recurrence0}
\end{equation}
for the first coefficient and
\begin{eqnarray}
&& (n+1)\,\textbf{c}_{n+1}(\vec{x},\vec{y}) - \Delta \textbf{c}_n(\vec{x},\vec{y}) + (\vec{x}-\vec{y})\cdot \vec{\nabla} \textbf{c}_{n+1}(\vec{x},\vec{y}) + \textbf{U}(\vec{x}) \,\, \textbf{c}_n(\vec{x},\vec{y}) + [\textbf{u}^2,\textbf{c}_n(\vec{x},\vec{y})] + \nonumber \\ && + \,\vec{\textbf{Q}}(\vec{x}) \cdot \vec{\nabla} \textbf{c}_n(\vec{x},\vec{y}) - \frac{1}{2} \, (\vec{x}-\vec{y}) \cdot \vec{\textbf{Q}}(\vec{x})\,\, \textbf{c}_{n+1} (\vec{x},\vec{y}) + 4 \pi \Big[ \Big(\delta_{n1}+\frac{1}{n!} \Big) \sum_{\ell=1}^{N_{\rm zm}}\,\Xi_{0\ell}(\vec{x}) \, \Xi_{0\ell}^\dagger (\vec{y}) + \label{recurrence1} \\ && + \,\frac{1}{n!}\, \sum_{\ell=1}^{N_{\rm zm}} (\vec{x}-\vec{y})\cdot \vec{\nabla} \Xi_{0\ell}(\vec{x}) \,\, \Xi_{0\ell}^\dagger (\vec{y})  - \frac{1}{2 \, (n!\,)} (\vec{x}-\vec{y})\cdot\vec{\textbf{Q}}(\vec{x}) \sum_{\ell=1}^{N_{\rm zm}} \Xi_{0\ell}(\vec{x}) \,\,\Xi_{0\ell}^\dagger (\vec{y}) \, \Big] \, \textbf{u}^{2n}=\textbf{0}\nonumber \, .
\end{eqnarray}
for the remaining ones. Note that the $n=0$ equation has been written separately because, given the choice of $\textbf{G}(\beta)$, zero modes do not enter at this order. Thus, the densities $\textbf{c}_{n}(\vec{x},\vec{y})$ for $n=1,2,3,\dots$ can be identified recursively using (\ref{recurrence0}) and (\ref{recurrence1}) in terms of the zero order density $\textbf{c}_{0}(\vec{x},\vec{y})$, which is fixed by the infinite temperature condition (\ref{initialconditionC}) and the definition (\ref{expansion5}), to be the constant $D\times D$
identity matrix: $\textbf{c}_0(\vec{x},\vec{y})=\textbf{I}_{D\times D}$.

Evaluation of the ${\cal H}$-spectral heat trace (\ref{heatfunction0}) requires to take the limit $\vec{y}\rightarrow \vec{x}$ of the densities before of integrating them. But sending the densities to the diagonal $\textbf{c}_n(\vec{x},\vec{x})$ and solving simultaneously the recurrence relations is a very subtle manoeuvre. The reason is that going to the $\vec{y}\rightarrow \vec{x}$ limit and computing partial derivatives with respect to $x_i$ as required in (\ref{recurrence0}) and (\ref{recurrence1}) are not mutually commuting operations. To handle this situation we introduce the $(\alpha_1,\alpha_2)$-order densities
\begin{equation}
{}^{(\alpha_1,\alpha_2)} \textbf{C}_n(\vec{x}) = \lim_{\vec{y}\rightarrow \vec{x}} \frac{\partial^{\alpha_1+\alpha_2} }{\partial x_1^{\alpha_1} \partial x_2^{\alpha_2}}\left(\textbf{c}_n(\vec{x},\vec{y})\right) \label{mayuscoef}
\end{equation}
where the partial derivatives are calculated first and the limit is taken later. Calculation of the partial derivative of the relations (\ref{recurrence0}) and (\ref{recurrence1}) of order $\alpha_1$ with respect to $x_1$ and order $\alpha_2$ with respect to $x_2$ and taking consecutively the limit $\vec{y}\rightarrow \vec{x}$ provide us with the recurrence relations for these diagonal magnitudes ${}^{(\alpha_1,\alpha_2)} \textbf{C}_n(\vec{x})$. The
partial derivatives of the first Seeley diagonal density ${}^{(\alpha_1,\alpha_2)} \textbf{C}_1(\vec{x})$ satisfy
\begin{eqnarray}
&& \hspace{-0.5cm} {}^{(\alpha_1,\alpha_2)} \textbf{C}_1(\vec{x}) = \frac{1}{\alpha_1+\alpha_2+1} \Big\{ {}^{(\alpha_1+2,\alpha_2)} \textbf{C}_0(\vec{x}) + {}^{(\alpha_1,\alpha_2+2)} \textbf{C}_0(\vec{x}) - \nonumber \\ && - \, \sum_{k_1=0}^{\alpha_1} \sum_{k_2=0}^{\alpha_2} {\alpha_1\choose k_1} {\alpha_2 \choose k_2} \frac{\partial^{k_1+k_2} \textbf{U}(\vec{x})}{\partial x_1^{k_1} \partial x_2^{k_2}} \,\,\,\, {}^{(\alpha_1-k_1,\alpha_2-k_2)} \textbf{C}_0(\vec{x}) \,\,-\,\, [\textbf{u}^2,{}^{(\alpha_1,\alpha_2)} \textbf{C}_0(\vec{x})] \,\,- \label{recurrence2} \\ && -
\sum_{k_1=0}^{\alpha_1} \sum_{k_2=0}^{\alpha_2} {\alpha_1 \choose k_1} {\alpha_2 \choose k_2} \Big[ \frac{\partial^{k_1+k_2} \textbf{Q}_1(\vec{x})}{\partial x_1^{k_1} \partial x_2^{k_2}}\,\,\,\, {}^{(\alpha_1-k_1+1,\alpha_2-k_2)} \textbf{C}_0(\vec{x}) \Big] + \nonumber \\ &&  - \sum_{k_1=0}^{\alpha_1} \sum_{k_2=0}^{\alpha_2} {\alpha_1 \choose k_1} {\alpha_2 \choose k_2} \Big[  \frac{\partial^{k_1+k_2} \textbf{Q}_2(\vec{x})}{\partial x_1^{k_1} \partial x_2^{k_2}} \,\,\,\,{}^{(\alpha_1-k_1,\alpha_2-k_2+1)} \textbf{C}_0(\vec{x})  \Big] + \nonumber \\ && + \frac{\alpha_1}{2} \sum_{k_1=0}^{\alpha_1-1} \sum_{k_2=0}^{\alpha_2} {\alpha_1-1 \choose k_1} {\alpha_2 \choose k_2} \frac{\partial^{k_1+k_2}\textbf{Q}_1(\vec{x})}{\partial x_1^{k_1} \partial x_2^{k_2}} \,\,\,\,{}^{(\alpha_1-k_1-1,\alpha_2-k_2)} \textbf{C}_1(\vec{x}) + \nonumber \\ && +  \frac{\alpha_2}{2} \sum_{k_1=0}^{\alpha_1} \sum_{k_2=0}^{\alpha_2-1} {\alpha_1 \choose k_1} {\alpha_2-1 \choose k_2} \frac{\partial^{k_1+k_2}\textbf{Q}_2(\vec{x})}{\partial x_1^{k_1} \partial x_2^{k_2}}\,\,\,\, {}^{(\alpha_1-k_1,\alpha_2-k_2-1)} \textbf{C}_1(\vec{x})  \Big\} \nonumber \nonumber
\end{eqnarray}
while the subsequent, $n>1$, derivatives of the diagonal Seeley densities verify the formula
{\small\begin{eqnarray}
&& \hspace{-0.5cm} {}^{(\alpha_1,\alpha_2)} \textbf{C}_{n+1}(\vec{x}) = \frac{1}{n+\alpha_1+\alpha_2+1} \Big\{ {}^{(\alpha_1+2,\alpha_2)} \textbf{C}_n(\vec{x}) + {}^{(\alpha_1,\alpha_2+2)} \textbf{C}_n(\vec{x}) - \nonumber\\ && - \sum_{k_1=0}^{\alpha_1} \sum_{k_2=0}^{\alpha_2} {\alpha_1\choose k_1} {\alpha_2 \choose k_2} \frac{\partial^{k_1+k_2} \textbf{U}(\vec{x})}{\partial x_1^{k_1} \partial x_2^{k_2}} \,\,\,\, {}^{(\alpha_1-k_1,\alpha_2-k_2)} \textbf{C}_n(\vec{x}) \,\, -\,\, [\textbf{u}^2,{}^{(\alpha_1,\alpha_2)} \textbf{C}_n(\vec{x})] \,\,- \nonumber \\ && -
\sum_{k_1=0}^{\alpha_1} \sum_{k_2=0}^{\alpha_2} {\alpha_1 \choose k_1} {\alpha_2 \choose k_2} \Big[ \frac{\partial^{k_1+k_2} \textbf{Q}_1(\vec{x})}{\partial x_1^{k_1} \partial x_2^{k_2}}\,\,\,\, {}^{(\alpha_1-k_1+1,\alpha_2-k_2)} \textbf{C}_n(\vec{x}) + \frac{\partial^{k_1+k_2} \textbf{Q}_2(\vec{x})}{\partial x_1^{k_1} \partial x_2^{k_2}} \,\,\,\, {}^{(\alpha_1-k_1,\alpha_2-k_2+1)} \textbf{C}_n(\vec{x})  \Big] + \nonumber \\ && + \,\, \frac{\alpha_1}{2} \sum_{k_1=0}^{\alpha_1-1} \sum_{k_2=0}^{\alpha_2} {\alpha_1-1 \choose k_1} {\alpha_2 \choose k_2} \frac{\partial^{k_1+k_2}\textbf{Q}_1(\vec{x})}{\partial x_1^{k_1} \partial x_2^{k_2}} \,\,\,\, {}^{(\alpha_1-k_1-1,\alpha_2-k_2)} \textbf{C}_{n+1}(\vec{x}) + \nonumber \\ && +  \,\, \frac{\alpha_2}{2} \sum_{k_1=0}^{\alpha_1} \sum_{k_2=0}^{\alpha_2-1} {\alpha_1 \choose k_1} {\alpha_2-1 \choose k_2} \frac{\partial^{k_1+k_2}\textbf{Q}_2(\vec{x})}{\partial x_1^{k_1} \partial x_2^{k_2}} \,\,\,\,{}^{(\alpha_1-k_1,\alpha_2-k_2-1)} \textbf{C}_{n+1}(\vec{x})-\label{recurrence3}  \\ && -\,\, 4\pi \Big( \delta_{n1}+\frac{\alpha_1+\alpha_2+1}{n!}\Big) \sum_{\ell=1}^{N_{\rm zm}} \frac{\partial^{\alpha_1+\alpha_2} \Xi_{0\ell}(\vec{x})}{\partial x_1^{\alpha_1} \partial x_2^{\alpha_2}} \,\,\,\Xi_{0\ell}^\dagger(\vec{x}) \,\, \textbf{u}^{2n} + \nonumber \\ && +\,\,\frac{2\pi\alpha_1}{n!} \sum_{k_1=0}^{\alpha_1-1} \sum_{k_2=0}^{\alpha_2} {\alpha_1-1\choose k_1} {\alpha_2 \choose k_2} \frac{\partial^{k_1+k_2} \textbf{Q}_1(\vec{x})}{\partial x_1^{k_1} \partial x_2^{k_2}} \sum_{\ell=1}^{N_{\rm zm}} \frac{\partial^{\alpha_1+\alpha_2-k_1-k_2-1}\Xi_{0\ell}(\vec{x})}{\partial x_1^{\alpha_1-k_1-1} \partial x_2^{\alpha_2-k_2}} \,\,\,\Xi_{0\ell}^\dagger (\vec{x}) \,\, \textbf{u}^{2n}\, + \nonumber \\ && +\,\,\frac{2\pi\alpha_2}{n!} \sum_{k_1=0}^{\alpha_1} \sum_{k_2=0}^{\alpha_2-1} {\alpha_1\choose k_1} {\alpha_2-1 \choose k_2} \frac{\partial^{k_1+k_2} \textbf{Q}_2(\vec{x})}{\partial x_1^{k_1} \partial x_2^{k_2}} \sum_{\ell=1}^{N_{\rm zm}} \frac{\partial^{\alpha_1+\alpha_2-k_1-k_2-1}\Xi_{0\ell}(\vec{x})}{\partial x_1^{\alpha_1-k_1} \partial x_2^{\alpha_2-k_2-1}}\,\,\, \Xi_{0\ell}^\dagger (\vec{x}) \,\, \textbf{u}^{2n} \,\,
\Big\} \nonumber
\end{eqnarray}}
Again, the choice of  $\textbf{G}(\vec{x})$ implies that derivatives of the first diagonal Seeley density ${}^{(\alpha_1,\alpha_2)} \textbf{C}_1(\vec{x})$ are not affected by the presence of the zero modes in the spectrum of ${\cal H}$. These recurrence relations start from the, constant, zero order Seeley densities:
\begin{equation}
{}^{(\alpha_1,\alpha_2)} \textbf{C}_0(\vec{x}) = \delta_{\alpha_1 0} \,\, \delta_{\alpha_2 0}\,\, \textbf{I}_{D\times D} \label{initialcondition3}
\end{equation}
which are directly identified from the infinite temperature condition and the definition (\ref{mayuscoef}). From (\ref{recurrence2}) and (\ref{recurrence3}) together with (\ref{initialcondition3}) we easily derive low order diagonal densities:
\begin{eqnarray}
{}^{(0,0)} \textbf{C}_1(\vec{x}) &=& - \textbf{U}(\vec{x}) \label{Seeley1}\\
{}^{(0,0)} \textbf{C}_2(\vec{x}) &=& -\frac{1}{6}\, \Delta \textbf{U}(\vec{x}) + \frac{1}{6} \, (\vec{\textbf{Q}}(\vec{x}) \cdot \vec{\nabla}) \, \textbf{U}(\vec{x}) + \frac{1}{12}  \, \vec{\textbf{Q}} (\vec{x}) \cdot \vec{\textbf{Q}}(\vec{x})  \,\, \textbf{U}(\vec{x}) -\frac{1}{6} \, (\vec{\nabla} \cdot \vec{\textbf{Q}}(\vec{x}))\, \, \textbf{U}(\vec{x}) +  \nonumber \\ &+& \frac{1}{2} \,\, \textbf{U}^2(\vec{x}) + \frac{1}{2}\,\, [\textbf{u}^2,\textbf{U}(\vec{x})] - 4 \pi \sum_{\ell=1}^{N_{\rm zm}} \Xi_{0\ell}(\vec{x}) \,\, \Xi_{0\ell}^\dagger (\vec{x}) \,\, \textbf{u}^2 \label{Seeley2} \, \, ,
\end{eqnarray}
where we observe that the impact of zero modes start at second order. In fact, all the new densities are the sum of the old Seeley densities plus terms induced by the zero modes proportional to $\textbf{u}^{2n-2}$. We remark that in this formula the vectorial notation $\vec{\textbf{Q}} (\vec{x}) \cdot \vec{\textbf{Q}}(\vec{x}) = \textbf{Q}_1(\vec{x})^2 + \textbf{Q}_2(\vec{x})^2$ has been used.
In the solution of the recurrence relations (\ref{recurrence2}) and (\ref{recurrence3}), e.g., up to order $n$, one needs to compute all the lower than $n$ densities and their derivatives. For instance, in order to obtain ${}^{(0,0)} C_6^{ab}(\vec{x})$ for ${\cal H}^+$ there is the need of knowing the diagonal densities and their derivatives ${}^{(\alpha_1,\alpha_2)} C_5^{ab}(\vec{x})$ for $\alpha_1,\alpha_2=0,1,2$ as data, which in turn demands the knowledge of ${}^{(\alpha_1,\alpha_2)} C_4^{ab}(\vec{x})$ for $\alpha_1,\alpha_2=0,1,2,3,4$, etcetera. It can be checked that the estimation of the Seeley densities ${}^{(0,0)} C_n^{ab}(\vec{x})$ at order $n$ demands the calculation of $\frac{8}{3}(n+1)(n+2)(4n+3)$ densities and their derivatives with lower $n$ of that type, a challenging task for a Mathematica program.

Formulas (\ref{factorizacion1}) and (\ref{expansion5}) allow us to write the diagonal of the heat integral kernel in the $\oplus_{a=1}^4 L_a^2(\mathbb{R}^2)$ Hilbert space as an asymptotic series in $\beta$:
\begin{equation}
\textbf{K}_{{\cal H}}(\vec{x},\vec{x};\beta) = \lim_{\vec{y}\rightarrow \vec{x}} \textbf{K}_{{\cal H}}(\vec{x},\vec{y};\beta) = \frac{1}{4\pi} \sum_{n=0}^\infty {}^{(0,0)}\textbf{C}_n(\vec{x}) \,\, \beta^{n-1}\,\, e^{-\beta \textbf{u}^2} + \sum_{\ell=1}^{N_{\rm zm}} \Xi_{0\ell}(\vec{x})\,\ \Xi_{0\ell}^\dagger (\vec{x}) \,\, \textbf{G} (\beta) \label{mgdwhk1}
\end{equation}
where, of course, ${}^{(0,0)}\textbf{C}_n(\vec{x}) = \textbf{c}_n(\vec{x},\vec{x})$ by definition. Spatial integration over $\mathbb{R}^2$ and taking the matrix trace of all the summands in (\ref{mgdwhk1})
offer us the asymptotic high temperature expansion
\begin{equation}
h_{{\cal H}}(\beta) - h_{{\cal H}_0} (\beta) = \frac{1}{4\pi} \sum_{n=1}^\infty \sum_{a=1}^D \, [\textbf{c}_n({\cal H}
)]_{aa} \,\,  e^{-\beta u_a^2} \,\, \beta^{n-1} + \sum_{\ell=1}^{N_{\rm zm}} \sum_{a=1}^D \, [f_\ell({\cal H})]_{aa} \,\, (1-e^{-\beta u_a^2}) \label{asymptoticseries0}
\end{equation}
for the difference between the spectral heat traces of the ${\cal H}$ and ${\cal H}_0$ operators. Here, we denote as
\begin{eqnarray*}
{} [c_n({\cal H})]_{aa} &=& \int_{\mathbb{R}^2} d^2 x \, \, [{}^{(0,0)}\textbf{C}_n(\vec{x})]_{aa} = \Big\langle [{}^{(0,0)}\textbf{C}_n(\vec{x})]_{aa} \Big\rangle \\
{} [f_\ell({\cal H})]_{aa} &=& \int_{\mathbb{R}^2} d^2 x \, \, (\Xi_{0\ell}(\vec{x}))_{a} (\Xi_{0\ell}^*(\vec{x}))_{a}= \Big\langle (\Xi_{0\ell}(\vec{x}))_{a}(\Xi_{0\ell}^*(\vec{x}))_{a} \Big\rangle \, \, \, ,
\end{eqnarray*}
the diagonal elements in the matrix sense of the Seeley coefficients, coming from integration of the diagonal elements in the functional sense of the Seeley densities.
The convention $\langle f(x) \rangle =\int_{\mathbb{R}^2} d^2 x \, f(\vec{x})$ will be used in some expressions later in the paper in order to alleviate the notation.

Another important spectral function is the generalized zeta function, formally defined as:
\begin{equation}
\zeta_{{\cal H}}(s) ={\rm Tr}_{L^2}\, {\cal H}^{-s} \, \,  \mbox{\lq\lq}="
 \sum_n \frac{1}{\omega_n^{2s}} \label{zetadef}
\end{equation}
which will play an essential r$\hat{\rm o}$le in the computation of the vortex mass quantum corrections. The spectral zeta function is a meromorphic function of the complex variable $s$ defined via analytic continuation following the Riemann zeta function pattern. Connection between the heat trace $h_{{\cal H}}(\beta)$ and the spectral zeta function $\zeta_{{\cal H}}(s)$ is established via Mellin transform,
\[
\zeta_{{\cal H}}(s) = \frac{1}{\Gamma[s]} \int_0^\infty d\beta \, \beta^{s-1} \, h_{{\cal H}} (\beta) \, \, .
\]
Application of this transformation to the asymptotic expansion (\ref{asymptoticseries0}) of the heat trace leads to the formula
\begin{equation}
\zeta_{{\cal H}}(s)-\zeta_{{\cal H}_0}(s) = \frac{1}{4\pi \,\Gamma[s]} \sum_{n=1}^\infty \sum_{a=1}^D \, [c_n({\cal H})]_{aa} \, (u_a^2)^{1-n-s} \, \Gamma[s+n-1] - \sum_{\ell=1}^{N_{\rm zm}} \sum_{a=1}^D \, [f_\ell({\cal H})]_{aa} \, (u_a^2)^{-s} \label{asymptoticseries1} \, \, \, ,
\end{equation}
which explicitly shows the meromorphic structure of this difference of generalized zeta functions with isolated poles located at the poles of the Euler Gamma function $\Gamma(s+n-1)$ and the singularities due to the zero modes regularized in the last term in (\ref{asymptoticseries1}). The residua at the poles are also easily identified.

\subsection{Spectral zeta function regularization of one-loop vortex mass shifts}

Standard lore in the semiclassical quantization of solitons tells us that the one-loop vortex mass shift $\Delta E_V$ in the AHM is the sum of two terms: (1) First, one computes the vortex Casimir energy, which is the energy of the state where all the vortex modes of fluctuation are unoccupied measured with respect to the energy of the state where the vacuum fluctuation modes are also unoccupied. {\footnote{This physical phenomenon is akin to the Casimir effect where the energy of photons in vacuum is subtracted from the energy of photons in presence of two conducting plates.}} (2) Second, the contribution of the mass renormalization counterterms up to one loop order is added in such a way that the remaining divergence in the Casimir energy, after subtraction of the zero point vacuum energy, is cancelled out. Identification of the mass renormalization counterterms in the Lagrangian is achieved in perturbation theory. Because we plan to renormalize particle masses we shall work the Feynman rules in the Feynman-'t Hooft renormalizable $R$-gauge. This gauge is the vacuum sector counterpart of the background gauge for fluctuations around the vortices. The $R$-gauge induces a complex ghost field $\chi(\vec{x},t)$ in the action functional  needed to restore the unitarity lost after adding the gauge fixing term. The ghost degrees of freedom give rise to its own Casimir energy and mass renormalization couterterms, which is subtracted -the ghosts are fermionic particles- to the corresponding energies coming from the bosonic field fluctuations. This routine is well established and standardized in the physical literature, see \cite{Alonso2004:prd,Alonso2005:prd,Alonso2006:hepth, Mateos2009:pos, Alonso2008:npb}. We shall denote the total contribution of the Casimir energies to the vortex classical energy as $\Delta E_V^C$, that of the mass renormalization counterterms as $\Delta E_V^R$, while the total vortex mass shift will be: $\Delta E_V= \Delta E_V^C + \Delta E_V^R$.

The self-dual vortex energies up to one-loop order in the AHM are the sum of the classical energies plus the energies of the fluctuations $\xi$. Choosing the background gauge and
accounting only the fluctuations at one-loop or quadratic order the vortex energy reads:
\[
E_V  =  \pi |n| v^2 + \frac{\hbar m}{2} \int d^2 x \,\, [\xi^T(\vec{x})  \,{\cal H}^+ \,\xi(\vec{x}) ] + o(\xi^3) \quad , \quad m=ev \, .
\]
The energy of ghosts, which is negative due to the fermionic character of these fictitious particles, is the sum of one quadratic and one interacting term:
\[
\Delta E_V^{\rm ghost}+ E_{\rm I}^{\rm ghost}=- \frac{\hbar m}{2} \int d^2 \vec{x} \,\, [ \,\chi^* (\vec{x})\, {\cal H}^G \,\chi(\vec{x})] - \frac{\hbar^2 e^2}{2}\int d^2 \vec{x}[ (\psi^*(\vec{x}) \varphi(\vec{x}) + \psi(\vec{x})\varphi^*(\vec{x})) \, \chi^*(\vec{x}) \chi(\vec{x})\,].
\]
The PDO operator ${\cal H}^G$ entering in the quadratic ghost term is
\[
{\cal H}^G = - \Delta + |\psi|^2 \, \, ,
\]
an ordinary Schr$\ddot{\rm o}$dinger operator that governs one-loop ghost fluctuations around the vortex, in contrast to ${\cal H}^+$ which is the matrix PDO (\ref{operator1}).

Thus, $\Delta E_V^C$ is the sum of the vortex Casimir energies of the bosonic $a_1,a_2,\varphi_1,\varphi_2$ fluctuations minus the Casimir energy of the fermionic fluctuation $\chi$. In sum, the vortex Casimir energy is given by the formal formula
\begin{equation}
\Delta E_V^C = \frac{\hbar m}{2} \Big[ \, {\rm Tr}_{\oplus_{a=1}^4 L^2_a(\mathbb{R}^2)} \, ({\cal H}^+)^\frac{1}{2} - {\rm Tr}_{\oplus_{a=1}^4 L^2_a(\mathbb{R}^2)} \, ({\cal H}_0)^\frac{1}{2} - [{\rm Tr}_{L^2(\mathbb{R}^2)}\,({\cal H}^{\rm G})^\frac{1}{2} - {\rm Tr}_{L^2(\mathbb{R}^2)}\,({\cal H}_0^{\rm G})^\frac{1}{2}\,] \, \Big] \label{Casimir1}
\end{equation}
where we recall that ${\cal H}_0 = -{\textbf I} \Delta + {\rm diag}\,(1,1,1,1)$ and ${\cal H}_0^G = -\Delta + 1$ are the corresponding second-order vacuum fluctuation operators.

The zeta function regularization procedure takes profit of the analytical continuation of the divergent quantity $\Delta E_V^C$ (\ref{Casimir1}) to the $s$-complex plane and assigning to the vortex Casimir energy its finite value at a regular point. This strategy is justified from the general theory about the analytical structure of spectral zeta functions of positive operators, in our problem we shall consider the spectral zeta functions of the PDO ${\cal H}^+$, ${\cal H}_0$, ${\cal H}^{\rm G}$, and ${\cal H}_0^{\rm G}$. Thus, we shall regularize the vortex Casimir energy in the form:
\begin{equation}
 \Delta E_V^C(s) = \frac{\hbar \mu}{2} \Big( \frac{\mu^2}{m^2}\Big)^s \Big\{ \zeta_{{\cal H}^+}(s) - \zeta_{{\cal H}_0}(s) - \Big( \zeta_{{\cal H}^G}(x) -\zeta_{{\cal H}_0^G}(s) \Big) \Big\} \, ,\label{Casimir0}
\end{equation}
where $\mu$ is a parameter of dimension $L^{-1}$ needed to keep correct the physical dimensions of energy away from the physical value $s=-\frac{1}{2}$: $ \Delta E_V^C= \lim_{s\rightarrow -\frac{1}{2}}  \Delta E_V^C(s)$.

The spectral heat kernel/zeta function control of divergences in QFT is a procedure that encompasses several different but related aspects.
\begin{enumerate}

\item Ultraviolet divergences arising in fluctuating topological defects are regularized by using the spectral zeta function of the Hessian operator. In odd dimensional spaces the zeta function giving the Casimir energy falls in a pole at $s=-\frac{1}{2}$ and one must go away from the pole in the $s$-complex plane to obtain a regularization of $\Delta E^C$, but in even dimensions the spectral zeta function is directly finite at the value of $s=-\frac{1}{2}$.

\item  The meromorphic structure of the spectral zeta function is clarified when it is obtained via Mellin transform of the heat kernel high temperature expansion. Poles appear in Euler Gamma functions $\Gamma(s+n-\frac{d}{2})$, i.e., at negative integers or zero values of $s+n-\frac{d}{2}$. Also, infrared divergences appear in the lower Seeley coefficients. Integration of low densities over the whole space gives rise to divergences proportional to the volume, or, the logarithm of the volume, etcetera. Regularization of these divergences requires to restrict the system to a cube of volume $V=l^d=(mL)^d$.

\item After these regularizations were performed some renormalizations have to be done. In $(1+1)$- or $(2+1)$-dimensional space-times,
where QFT models are usually superrenormalizable, zero point and mass renormalization, taming the divergences due to the tadpoles and self-energy graphs, are enough.

\item It remains to deal with the delicate question of finite renormalizations. We shall stick to the heat kernel renormalization criterion,
tantamount to the vanishing of the tadpole graph. In the limit of infinite particle masses there are no quantum fluctuations, thus there should be no quantum corrections. This means that the contribution of all the coefficients multiplied by non negative powers of mass must be
exactly cancelled in the renormalization process. In one and two spatial dimensions only $\textbf{c}_0$ and $\textbf{c}_1$ survive when the particles become infinitely heavy and the annihilation of their contribution fixes our renormalization criterion.

\item Zero modes, however, respond to rigid motions which survive in the infinite mass regime and the above criterion does not apply to their contributions.

\end{enumerate}

The heat kernel/zeta function technology applied in the computation of (\ref{Casimir1}) requires to write (\ref{asymptoticseries1}) for both the ${\cal H}={\cal H}^+$ and ${\cal H}={\cal H}^G$ operators arising in the Abelian Higgs model. The difference between the spectral zeta functions of the PDO's ${\cal H}^+$ and ${\cal H}_0$ reads:
\[
\zeta_{{\cal H}^+}(s) - \zeta_{{\cal H}_0}(s) = \frac{1}{4\pi \,\Gamma[s]} \sum_{n=1}^\infty \sum_{a=1}^4 \, \frac{[c_n({\cal H}^+)]_{aa}}{u^{2n+2s-2}} \, \Gamma[s+n-1] - \sum_{\ell=1}^{N_{\rm zm}} \sum_{a=1}^4 \, [f_\ell({\cal H}^+)]_{aa} u^{-2s} \,
\]
where, although the $\vert\vec{x}\vert\to +\infty$ asymptotics of the matrix potential in ${\cal H}^+$ is $\textbf{u}={\rm diag}\,\{1,1,1,1\}$, we have written $u_a=u$, $a=1,2,3,4$, in order to later analyze the $u\to +\infty$, infinite particle masses, limit. We remark that the subtraction of $\zeta_{{\cal H}_0}(s)$ corresponds exactly to zero point renormalization:
\[
\lim_{s\to -\frac{1}{2}}\, \frac{1}{4\pi}\sum_{a=1}^4\, \frac{c_0[{\cal H}^+]_{aa}}{u^{2s-2}}\Gamma[s-1]=u^3\frac{l^2}{\pi}\Gamma[-{\textstyle\frac{3}{2}}]=\zeta_{{\cal H}_0}(-{\textstyle\frac{1}{2}}) \, \, .
\]
The lower Seeley coefficients are easily obtained from (\ref{Seeley1}) and (\ref{Seeley2}):
\begin{eqnarray*}
\sum_{a=1}^4 [c_1({\cal H}^+)]_{aa} &=& \Big\langle 5(1-|\psi|^2) - 2 V_kV_k  \Big\rangle \\
\sum_{a=1}^4 [c_2({\cal H}^+)]_{aa} &=& \Big\langle -\frac{5}{6} \Delta |\psi|^2 -\frac{1}{3} \Delta(V_kV_k) + 4 \sum_{i,j=1}^2 (D_i \psi_j)^2 + \frac{13}{4} (1-|\psi|^2)^2 -\\ && -2V_kV_k(1-|\psi|^2) + \frac{1}{3} (V_kV_k)^2 \Big\rangle -4 \pi \sum_{\ell=1}^{N_{\rm zm}} \sum_{a=1}^4 \, [f_\ell({\cal H}^+)]_{aa}u^2 \, \, ,
\end{eqnarray*}
where we observe also that the new Seeley coefficients are the sum of the old coefficients plus the last term induced by the zero modes.

Simili modo, the ghost spectral zeta function regularizes the ghost Casimir energy:
\[
\zeta_{{\cal H}^{\rm G}}(s) - \zeta_{{\cal H}_0^{\rm G}}(s) = \frac{1}{4\pi \,\Gamma[s]} \sum_{n=1}^\infty \, \frac{c_n({\cal H}^{\rm G})}{u^{2s+2n-2}} \, \Gamma[s+n-1] - \sum_{\ell=1}^{N_{\rm zm}^G} \, f_\ell({\cal H}^{\rm G}) \, u^{-2s}
\]
Again, we leave free the asymptotic value  of $U^{\rm G}(\vec{x})$ to ponder the heat kernel renormalization criterion, although we know that $U^{\rm G}(\vec{x})=\vert\psi \vert^2(\vec{x})\, \, \,  \equiv \, \, \, u=1$ for the vortex. $N_{\rm zm}^{\rm G}$ denotes the zero mode number in the ${\cal H}^{\rm G}$-spectrum. The first and second ghost Seeley coefficients are:
\[
c_1( {\cal H} ^{\rm G}) = \left\langle 1- |\psi|^2 \right\rangle \hspace{0.5cm}\mbox{and}\hspace{0.5cm} c_2( {\cal H} ^{\rm G}) = \left\langle -{\textstyle\frac{1}{6}} \Delta |\psi|^2 + {\textstyle\frac{1}{2}}(|\psi|^2-1)^2 \right\rangle -4 \pi \sum_{\ell=1}^{N_{\rm zm}^G} \, f_\ell({\cal H}^{\rm G}) u^2
\]
Because zero modes $\Xi_{0\ell}(\vec{x})$ are orthogonal to each other and normalized it is clear that:
\[
\sum_{\ell=1}^{N_{\rm zm}} \sum_{a=1}^4 \, [f_\ell({\cal H}^+)]_{aa} = N_{\rm zm} \hspace{0.5cm} \mbox{and} \hspace{0.5cm} \sum_{\ell=1}^{N_{\rm zm}^G} \, f_\ell({\cal H}^{\rm G}) = N_{\rm zm}^{\rm G} \, .
\]
But the vortex zero mode number is $2N$, twice the vorticity, and $0$ for the ghost fluctuation operator ${\cal H}^{\rm G}$ which is a positive operator. Thus, $N_{\rm zm}=2N$ and $N_{\rm zm}^{\rm G}=0$ and the total BPS vortex Casimir energy (\ref{Casimir0}) is
\begin{equation}
\lim_{s\rightarrow -\frac{1}{2}}  \Delta E_V^C(s) = \lim_{s\rightarrow -\frac{1}{2}}  \frac{\hbar \mu}{2} \Big( \frac{\mu^2}{m^2}\Big)^s \Big\{ \frac{1}{4\pi \,\Gamma[s]} \sum_{n=1}^\infty \Big( \sum_{a=1}^4\frac{[\textbf{c}_n({\cal H}^+)]_{aa}}{u^{2s+2n-2}} - \frac{c_n({\cal H}^G)}{u^{2n+2s-2}} \Big) \, \Gamma[s+n-1] -2N u^{-2s}\, \Big\} \, \, \, . \label{vortcas}
\end{equation}
Note that the first summand in (\ref{vortcas}) is proportional to $u$:
\[
\frac{\hbar m}{2}\frac{1}{4\pi\Gamma(-\frac{1}{2})}\Big(\sum_{a=1}^4\, c_1({\cal H}^+)_{aa}-c_1({\cal H}^{\rm G})\Big)\Gamma(-{\textstyle\frac{1}{2}}) u=
\frac{\hbar m}{4\pi}\cdot \Big\langle 2(1-\vert\psi\vert^2)-V_kV_k\Big\rangle \cdot u \, .
\]
Therefore the contribution of this term must be exactly annihilated in a renormalization procedure adjusted to suppress it without leaving any finite remnants. The next term is proportional to $1/u$ and, thus, is susceptible to be kept, as well as all the higher order than $2$ terms.

In fact, the only renormalization, after control of the zero point divergences, remaining in the planar AHM is the mass renormalization. In References \cite{Alonso2004:prd} and \cite{Alonso2006:hepth}, together with other collaborators, we identified the energy induced by the counterterms needed to
tame the tadpoles and self-energy divergent graphs in a minimal renormalization scheme, i.e., only subtracting the infinities arising in these graphs. The divergent mass renormalization energy is:
\[
\Delta E_V^R =  2 \,\hbar\, m \,I(u) \, \Big\langle\Sigma_1(\psi,V_k)\Big\rangle
\]
where
\[
\Big\langle\Sigma_1(\psi,V_k)\Big\rangle=\int_{\mathbb{R}^2} d^2 x \Big[ 1-|\psi|^2 - \frac{1}{2} V_kV_k \Big] = \Big\langle 1-|\psi|^2 - \frac{1}{2} V_kV_k \Big\rangle \quad ,
\]
obviously proportional to $\sum_{a=1}^4\, c_1({\cal H}^+)_{aa}-c_1({\cal H}^{\rm G})$, and $I(u)$ is the divergent integral
\[
I(u)=\frac{1}{2} \int_{-\infty}^\infty \frac{d^2k}{(2\pi)^2} \frac{1}{\sqrt{k_1^2+k_2^2+u^2}}
\]
arising in closed loop propagators. The idea is to regularize also $I(u)$ by means of the zeta function procedure:
\[
I(u,s)=\frac{1}{2} \int_{-\infty}^\infty \frac{d^2k}{(2\pi)^2} \frac{1}{(k_1^2+k_2^2+u^2)^{s+1}} = \frac{1}{2}\zeta_{-\Delta +u^2} (s+1) \, \, ,
\]
which implies that:
\[
I(u)=I(u,-{\textstyle\frac{1}{2}})= \lim_{s\to -\frac{1}{2}} \, \zeta_{-\Delta +u^2} (s+1)=  \zeta_{-\Delta +u^2}({\textstyle\frac{1}{2}})\, .
\]
Recall that ${\cal H}_0$ is a $4\times 4$ diagonal matrix PDO whose components are Helmoltz operators: $-\Delta+u^2$. Thus, $\zeta_{-\Delta+u^2}(\frac{1}{2})= \frac{1}{4} \,\zeta_{{\cal H}_0}(\frac{1}{2})$. Moreover, we knew that,
\[
\zeta_{{\cal H}_0}(s)=\frac{1}{\pi} \frac{\Gamma[s-1]}{\Gamma[s]} u^{2-2s} \hspace{1cm} \mbox{and} \hspace{1cm} \zeta_{{\cal H}_0^{\rm G}}(s)=\frac{1}{4\pi} \frac{\Gamma[s-1]}{\Gamma[s]} u^{2-2s} \, \, ,
\]
therefore, the regularized mass renormalization energy reads:
\[
\Delta E_V^R = \lim_{s\rightarrow -\frac{1}{2}} \Delta E_V^R(s)= \lim_{s\rightarrow -\frac{1}{2}}  \frac{\hbar \mu}{4 \pi} \Big( \frac{\mu^2}{m^2} \Big)^s \frac{\Gamma[s]}{\Gamma[s+1]} \cdot\, u^{-2s} \cdot\, \Big\langle\Sigma_1(\psi, V_k)\Big\rangle \, .
\]
The sum of the analytical continuations of the Casimir and mass renormalization energies $\Delta E_V^C(s) + \Delta E_V^R(s)$ is:
\begin{eqnarray*}
&& \hspace{-0.7cm}\Delta E_V^C(s) + \Delta E_V^R(s)= \frac{\hbar \mu}{2} \Big( \frac{\mu^2}{m^2}\Big)^s \Big\{ \frac{1}{4\pi} \Big\langle 5(1-|\psi|^2)- 2 V_kV_k\Big\rangle u^{-2 s}-  \frac{1}{4\pi} \left\langle 1-|\psi|^2 \right\rangle u^{-2s}\\ && +\frac{1}{4\pi \Gamma[s]} \sum_{n=2}^\infty \sum_{a=1}^4 \frac{[\textbf{c}_n({\cal H}^+)]_{aa}}{u^{2n+2s-2}} \Gamma[s+n-1] -2N u^{-2s} - \frac{1}{4\pi \Gamma[s]} \sum_{n=2}^\infty \frac{c_n({\cal H}^G)}{u^{2s+2n-2}} \Gamma[s+n-1] + \\ && + \frac{1}{2\pi} \frac{1}{s} \left\langle 1-|\psi|^2 - \frac{1}{2} V_kV_k \right\rangle \cdot u^{-2s}
  \Big\} = \frac{\hbar \mu}{2} \Big( \frac{\mu^2}{m^2}\Big)^s \Big\{\Big( \frac{1}{\pi} + \frac{1}{2\pi s} \Big) \left\langle 1- |\psi|^2 - \frac{1}{2} V_kV_k \right\rangle \cdot u^{-2s} +  \\
&& + \frac{1}{4\pi \Gamma[s]} \sum_{n=2}^\infty \Big( \sum_{a=1}^4 \frac{[c_n({\cal H}^+)]_{aa}}{u^{2s+2n-2}} - \frac{c_n({\cal H}^G)}{u^{2s+2n-2}} \Big) \Gamma[s+n-1] - 2N u^{-2s} \Big\} \, \, ,
\end{eqnarray*}
where we have used $\Gamma(s+1)=s\Gamma(s)$. The key observation is that, according to the heat kernel renormalization criterion, the contribution of the first order Seeley coefficients is exactly cancelled by the
minimal subtraction scheme chosen in our mass renormalization prescription. This statement can be easily checked by looking at the first term in the last equality at the physical value $s=-\frac{1}{2}$.
Therefore, the one-loop BPS vortex mass shift is obtained in this approach by the asymptotic formula:
\[
\Delta E_V =  \lim_{s\rightarrow -\frac{1}{2}} [\Delta E_V^C(s) + \Delta E_V^R(s)]
\]
which provides us with the final response
\begin{equation}
\Delta E_V = -\frac{\hbar m}{16 \pi^\frac{3}{2}} \sum_{n=2}^\infty \Big( \sum_{a=1}^4 [\textbf{c}_n({\cal H}^+)]_{aa} - c_n({\cal H}^G) \Big) \, \Gamma[n-{\textstyle\frac{3}{2}}] - \hbar m N \label{quantumshift} \, \, .
\end{equation}

\subsection{One-loop mass shifts of BPS rotationally symmetric vortices: surge of weak quantum forces}

Use of formula (\ref{quantumshift}) guides us towards the computation of one-loop mass shifts for BPS circularly symmetric vortices, solutions of the PDE system (\ref{pde1}) of the form
\[
\psi(x_1,x_2)=f_N(r) \, e^{iN\theta} \hspace{0.5cm};\hspace{0.5cm} V_r(r,\theta)=0 \, \, \, , \, \, \, V_\theta(r,\theta)= \frac{N}{r} \, \beta_N(r)
\]
where $r=\sqrt{x^1x^1+x^2x^2}$ and $\theta={\rm arctan}\frac{x^2}{x^1}$ are polar coordinates in the plane. In this case the just mentioned PDE system becomes the ODE system:
\begin{equation}
f_N'(r)=\frac{N}{r} f_N(r) [1-\beta_N(r)]\hspace{0.5cm};\hspace{0.5cm} \beta_N'(r)=\frac{r}{2N}[1-f_N^2(r)] \label{ode1} \, \, .
\end{equation}
The subindex $N$ in $f_N(r)$ and $\beta_N(r)$ reminds us that the radial profiles depend on the vorticity $N$, i.e., they are different in different topological sectors. The well known procedure for finding solutions of these ordinary equations proceed in three steps: (1) Solving the (\ref{ode1}) near $r=0$ one finds $f_N(r)\simeq_{r\to 0} D_N r^N$ and $\beta_N(r)\simeq_{r\to 0} E_N r^2$, where $D_N$ and $E_N$ are integration constants, that are regular solutions near the origin. (2) The asymptotic conditions (\ref{asymptotic}) demand that $f_N(r)\rightarrow 1$ and $\beta_N(r)\rightarrow 1$ in the $r\rightarrow \infty$ limit. One solves then the (\ref{ode1}) system very far from the origin. An smooth sewing between the two regimes requires a precise choice of $D_N$ and $E_N$. (3) This shooting procedure is  numerically implemented to build interpolating solutions to (\ref{ode1}) at intermediate distances. In this way the circularly symmetric BPS $N$-vortex solutions are obtained, and these $N$-vortex profiles are basic ingredients in the one-loop BPS vortex mass shift formula (\ref{quantumshift}).

The remaining ingredients needed in formula (\ref{quantumshift}) are the $2N$ orthonormal zero mode fluctuations of the circularly symmetric $N$-vortices of the form, see \cite{Alonso2016:jhep},
{\small\begin{equation}
\xi_0(\vec{x},N,k)= r^{N-k-1} \left( \begin{array}{c} h_N(r) \, \sin[(N-k-1)\theta] \\ h_N(r) \, \cos[(N-k-1)\theta] \\ - \frac{h_N'(r)}{f_N(r)} \, \cos(k\theta) \\ - \frac{h_N'(r)}{f_N(r)} \, \sin(k\theta) \end{array} \right)
\,
 , \, \, \, \xi_0^\perp(\vec{x},N,k)= r^{N-k-1} \left( \begin{array}{c} h_N(r) \, \cos[(N-k-1)\theta] \\ -h_N(r) \, \sin[(N-k-1)\theta] \\  - \frac{h_N'(r)}{f_N(r)} \, \sin(k\theta) \\  \frac{h_N'(r)}{f_N(r)} \, \cos(k\theta)  \end{array} \right) \, , \label{zeromode4}
\end{equation}}
where $k=0,1,2,\dots,N-1$, and the zero mode radial profile $h_N(r)$ verifies the ODE
\begin{equation}
-r \, h_N''(r)+[1+2k-2N\,\beta_N(r)]\,h_N'(r) + r \,f_N^2(r)\, h_N(r)=0 \label{ode5}
\end{equation}
with boundary conditions $h_N(0)\neq 0$ and $\lim_{r\rightarrow \infty} h_N(r)=0$. Again a numerical approach applied to solve (\ref{ode5}) with the just prescribed conditions at the origin and at infinity offer us quite precise knowledge of the $2N$ zero mode fluctuations of a BPS vortex solution with vorticity $N$ \cite{Alonso2016:jhep}. All this information allows us to use the recurrence relations (\ref{recurrence2}) and (\ref{recurrence3}) in order to obtain the Seeley coefficients $\sum_{a=1}^4 [\textbf{c}_k({\cal H}^+)]_{aa}$ and $c_k({\cal H}^G)$ entering in the vortex mass quantum correction formula (\ref{quantumshift}). The practical use of (\ref{quantumshift}) involves the truncation of the series at a finite order $n_T$, i.e., replacing the series by the partial sum:
\begin{equation}
\Delta E_V = -\frac{\hbar m}{16 \pi^\frac{3}{2}} \sum_{n=2}^{n_T} \Big( \sum_{a=1}^4 [\textbf{c}_n({\cal H}^+)]_{aa} - c_n({\cal H}^G) \Big) \, \Gamma[n-{\textstyle\frac{3}{2}}] - \hbar m N \label{truncaquantumshift} \, \, .
\end{equation}
We estimate the vortex mass quantum correction by applying (\ref{truncaquantumshift}) with $n_T=6$. Computation of the lower six Seeley coefficients requires the calculation of 4043 functional coefficients ${}^{(\alpha,\gamma)}C_n^{ab}(\vec{x})$. We develope this program by using the symbolic software platform \textit{Mathematica}. The code of this task can be found in the web page http//:campus.usal.es/$\sim$mpg/General/MathematicaTools.htm. Estimation of the matrix and functional traces of these densities provides us with the previously mentioned Seeley coefficients. The results are displayed in Table 1 and Table 2.

\begin{table}[h]
\centering\begin{tabular}{|c||c|c|c|c|c|} \hline
    & \multicolumn{5}{|c|}{${\rm tr}([\textbf{c}_n({\cal H}^+)])$}  \\ \hline
$n$ & $N=1$ & $N=2$  & $N=3$ & $N=4$ & $N=5$  \\ \hline \hline
$2$ & $5.20990655$ & $10.75849898$ & $14.59990245$ & $17.58450757$ & $20.05604942$ \\
$3$ & $0.60457807$ & $0.64034809$  & $-1.43031758$ & $-5.93852744$ & $-13.0290730$  \\
$4$ & $0.10055209$ & $-0.23427492$ & $-1.42368210$ & $-3.57770210$ & $-6.70544685$ \\
$5$ & $0.02634327$ & $-0.11250983$ & $-0.50804216$ & $-1.20295070$ & $-2.21121000$  \\
$6$ & $0.00468414$ & $-0.03251509$ & $-0.12931186$ & $-0.29574333$ & $-0.53589538$  \\ \hline
\end{tabular}

\caption{\it Values of the $n$-th Seeley coefficients for the small $N$-vortex fluctuation operator ${\cal H}^+$ entering in the planar vortex mass quantum correction (\ref{quantumshift}).}
\end{table}

\begin{table}[h]
\centering\begin{tabular}{|c||c|c|c|c|c|} \hline
    & \multicolumn{5}{|c|}{${\rm tr}([\textbf{c}_n({\cal H}^G)])$}  \\ \hline
$n$ & $N=1$ & $N=2$  & $N=3$ & $N=4$ & $N=5$  \\ \hline \hline
$2$ & $2.60573638$ & $6.80907379$ & $11.49149074$ & $16.45567556$ & $21.55628055$ \\
$3$ & $0.31910464$ & $1.34189515$ & $2.60714103$  & $4.00530969$ &  $5.48466835$ \\
$4$ & $0.02297681$ & $0.20498547$ & $0.46776735$  & $0.77192241$ &  $1.10205597$ \\
$5$ & $0.00122645$ & $0.02380029$ & $0.06735758$  & $0.12074591$ &  $0.18031589$ \\
$6$ & $0.00006965$ & $0.00219104$ & $0.00800451$  & $0.01580181$ &  $0.02478549$ \\ \hline
\end{tabular}

\caption{\it Values of the $n$-th Seeley coefficients for the ghost operator ${\cal H}^G$ entering in the planar vortex mass quantum correction (\ref{quantumshift}).}
\end{table}

In Table 3 we display the response of this formula up to $n_T=6$. The last row offers the best estimation of the BPS $N$-vortex mass quantum correction. In the graphic we observe that the mass shift of a circularly symmetric vortex of vorticity $N$ is greater (less negative) than the mass shift of $N$ quanta of magnetic flux infinitely appart from each other. This means that one-loop fluctuations induce (very weak) repulsive forces between vortices, or, equivalently, that BPS vortices are pushed by quantum fluctuations towards a Type II superconductivity phase.

\begin{table}[h]
\centering\begin{tabular}{|c|c|c|c|c|c|} \hline
$n_T$ & $\rule[-0.3cm]{0cm}{0.9cm} \frac{\Delta E_V^{N=1}}{\hbar m}$ & $\frac{\Delta E_V^{N=2}}{\hbar m}$ & $\frac{\Delta E_V^{N=3}}{\hbar m}$ & $\frac{\Delta E_V^{N=4}}{\hbar m}$ & $\frac{\Delta E_V^{N=5}}{\hbar m}$ \\ \hline\hline
$2$ & $-1.0518$ & $-2.0786$ & $-3.0618$ & $-4.0225$ & $-4.9701$ \\
$3$ & $-1.0546$ & $-2.0716$ & $-3.0217$ & $-3.9235$ & $-4.7860$ \\
$4$ & $-1.0558$ & $-2.0650$ & $-2.9935$ & $-3.8586$ & $-4.6695$ \\
$5$ & $-1.0567$ & $-2.0599$ & $-2.9720$ & $-3.8093$ & $-4.5803$ \\
$6$ & $-1.0573$ & $-2.0554$ & $-2.9541$ & $-3.7686$ & $-4.5071$ \\ \hline
\end{tabular}\hspace{1cm}\begin{tabular}{c} \includegraphics[height=3cm]{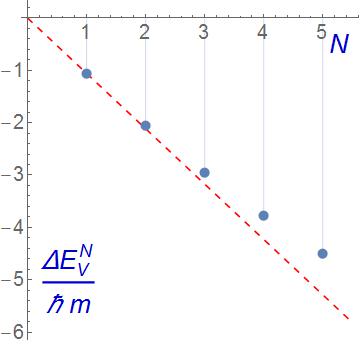}\end{tabular}
\caption{\it Estimation of the quantum correction to the $N$-vortex mass up to vorticity $N=5$ computed from the $2\leq n_T \leq 6$ partial sums of the series (\ref{quantumshift}).}
\end{table}

\section{One-loop string tension shifts for cylindrically symmetric BPS vortex filaments}

In this Section we shall try to compute one-loop BPS vortex tension shifts in the (3+1)-dimensional AHM. The BPS planar vortex solutions assuming cylindrical symmetry, i.e., infinitely repeated in the new dimension, become
the famous self-dual Abrikosov-Nielsen-Olesen magnetic filaments or tubes. The AHM action functional in $(3+1)$ Minkowski space-time at the BPS point is:
\[
S[\phi,A]=\frac{1}{e^2}\int d^4 x \Big[ - \frac{1}{4} F_{\mu \nu} F^{\mu \nu} + \frac{1}{2} (D_\mu \phi)^* D^\mu \phi -\frac{1}{8} (\phi^*\phi-1)^2 \Big] \, \, .
\]
The differences with respect to the planar AHM action are: (1) $d^4x=dx^0dx^1dx^2dx^3$; (2) $\vec{x}=x^1\vec{e}_1+x^2\vec{e}_2+x^3\vec{e}_3$ where $\vec{e}_i\cdot\vec{e}_j=\delta_{ij}$, $i,j=1,2,3$; (3) $g_{\mu\nu}={\rm diag}(1,-1,-1,-1)$ with $\mu, \nu=0,1,2,3$; (4) the gauge connection has four components: $A_\mu=(A_0,A_1,A_2,A_3)$ and (5) the antisymmetric EM tensor field $F_{\mu\nu}=\frac{\partial A_\nu}{\partial x^\mu} -\frac{\partial A_\mu}{\partial x^\nu}$ encompasses $6$ independent components: $3$ components of the electric field $E_i(\vec{x})= F_{0i}(\vec{x})$ and $3$ components of the magnetic field $B_i(\vec{x})=\frac{1}{2}\varepsilon_{ijk} F_{jk}(\vec{x})$.

Static cylindrically symmetric field configurations are independent of $x^0$ and $x^3$: $\phi=\phi(x_1,x_2)$, $A_\alpha=A_\alpha(x_1,x_2)$. To make this restriction gauge invariant we choose the temporal and axial gauges: $A_0=A_3=0$. For configurations with these symmetries the first-order PDE (\ref{ode1}) system still admit BPS vortex solutions. Seen in three dimensions, the BPS vortices become cylindrical magnetic tubes with the axis along the third dimension $x^3$ and, therefore the planar solutions are the cross sections at $x^3$ fixed of these stringy topological defects. Like in the previous Sections, we are interested in studying the one-loop fluctuations around these infinitely long BPS vortex filaments. The main novelty here are the fluctuations in the third dimension, i.e., the fluctuations are functions also of $x^3$: $\varphi(x^1,x^2,x^3)$. Moreover, although the axial gauge has been chosen to fix the BPS vortex solutions, perturbations in the third component of the gauge potential must be taken into account:
\[
\phi(\vec{x})=\psi(x^1,x^2)+\varphi(\vec{x}) \qquad , \qquad A_\alpha(\vec{x})=V_\alpha(x^1,x^2)+a_\alpha(\vec{x}) \, \, \, , \alpha=1,2 \, \, \, , \, \, \,  A_3(\vec{x})=a_3(\vec{x})
\]
The vortex filament fluctuations are assembled in a five  component column vector $\xi(\vec{x})$ that includes also fluctuations in the third component of the vector potential $a_3(\vec{x})$:
\[
\xi(x^1,x^2,x^3)=\left( \begin{array}{ccccc} a_1(x^1,x^2,x^3) & a_2(x^1,x^2,x^3) & a_3(x^1,x^2,x^3) & \varphi_1(x^1,x^2,x^3) & \varphi_2(x^1,x^2,x^3) \end{array} \right)^t \, .
\]
To exclude spurious pure gage fluctuations we impose the background gauge
\[
B(a_k,\varphi,\phi)=\sum_{j=1}^3 \partial_j a_j(\vec{x}) -\left[ \psi_1(x^1,x^2) \varphi_2(\vec{x}) - \psi_2(x^1,x^2) \varphi_1(\vec{x}) \right]=0
\]
Expanding the classical action plus the gauge fixing term up to the quadratic order in $\xi$ we unveil the second-order fluctuation operator:
{\small\begin{equation}
{\cal L}= \left( \begin{array}{ccccc}
-\Delta + |\psi|^2 & 0 & 0 & -2D_1 \psi_2 & 2 D_1 \psi_1 \\
0 & -\Delta +|\psi|^2 & 0 & -2 D_2 \psi_2 & 2 D_2 \psi_1 \\
0 & 0 & - \Delta +  |\psi^2| & 0 & 0\\
-2 D_1 \psi_2 & -2 D_2\psi_2 & 0 & -\Delta +\frac{1}{2} (3|\psi|^2-1)+V_k V_k & -2 V_k \partial_k -\partial_k V_k \\
2D_1\psi_1 & 2 D_2 \psi_1 & 0 &2V_k \partial_k + \partial_k V_k & -\Delta +\frac{1}{2} (3|\psi|^2-1) + V_k V_k
 \end{array} \right) \label{operatorL}
\end{equation}}
We remark that in $3$D the three-dimensional Laplacian enters : $\Delta=\frac{\partial^2}{\partial x_1^2} + \frac{\partial^2}{\partial x_2^2} + \frac{\partial^2}{\partial x_3^2}$. The structure of the matrix PDO (\ref{operatorL}) shows that the $a_3$-fluctuations are decoupled and do not mix with the other four fluctuations. Therefore, one-loop string tension shifts to be extracted from the spectrum  of ${\cal L}$-fluctuations come from the spectra of the two operators
\[
{\cal K}= - \mathbf{I}_{4\times 4} \frac{\partial^2}{\partial x_3^2} + {\cal H}^+ \hspace{0.8cm},\hspace{0.8cm} {\cal L}_3= - \Delta +  |\psi^2|
\]
accounted for separately. The matrix PDO ${\cal K}$ is in turn obtained by adding to the 1D Laplacian along the $x_3$-axis times the $4\times 4$ unit matrix the old Hessian operator (\ref{operator1}) working in the $(2+1)$D  AHM, fully analyzed in previous Sections. It is clear that the eigenvalues of the ${\cal K}$ operator, ${\cal K} F_n(\vec{x})= \varepsilon_n^2 F_n(\vec{x})$ are of the form
\[
\varepsilon_n^2 = \omega_n^2 + k_3^2
\]
where $\omega_n^2$ are the eigenvalues of ${\cal H}^+$. $k_3\in \mathbb{R}$ belongs to the continuous spectrum of the
$1$D Laplacian and has spectral density $\rho(k_3)=\frac{l}{2\pi}$ when particle motion in the third spatial dimension $x_3$ is confined to an interval of (non dimensional) length $2l=2 m L$, which eventually will go to infinity. The ${\cal K}$-heat function $H_{\cal K}(\beta)$, after subtraction of the ${\cal K}_0$-heat function where ${\cal K}_0$ is obtained by replacing ${\cal H}^+$ with ${\cal H}_0$, is essentially obtained from $h_{{\cal H}^+}(\beta)$ and $h_{{\cal H}_0}(\beta)$:
\begin{eqnarray*}
H_{\cal K}(\beta)- H_{{\cal K}_0}(\beta) &=& {\rm Tr}_{{\rm L}^2}\, e^{-\beta {\cal K}}-  {\rm Tr}_{{\rm L}^2}\, e^{-\beta {\cal K}_0} = \int_{-\infty}^\infty dk_3 \frac{l}{2\pi} \Big[ h_{{\cal H}^+}(\beta) - h_{{\cal H}_0}(\beta) \Big] e^{-\beta k_3^2} = \\&=& \frac{l}{2\sqrt{\pi}} \beta^{-\frac{1}{2}} [h_{\cal H}(\beta) - h_{{\cal H}_0}(\beta) ] \, \, .
\end{eqnarray*}
The Mellin transform allows us to calculate the difference between the spectral zeta functions ${\cal Z}_{\cal K}(s)- {\cal Z}_{{\cal K}_0}(s)$ of the ${\cal K}$ and ${\cal K}_0$ operators:
\begin{eqnarray*}
{\cal Z}_{\cal K}(s)- {\cal Z}_{{\cal K}_0}(s) &=& \frac{1}{\Gamma[s]} \int_0^\infty d\beta \beta^{s-1} [H_{\cal K}(\beta)-H_{{\cal K}_0}(\beta)] = \frac{1}{\Gamma[s]} \int_0^\infty d\beta \frac{l}{2\sqrt{\pi}} \beta^{s-\frac{3}{2}} [h_{\cal H}(\beta) - h_{{\cal H}_0}(\beta)]= \\ &=& \frac{1}{\Gamma[s]} \frac{ l}{2\sqrt{\pi}} \Gamma[s-{\textstyle\frac{1}{2}}] \Big[ \zeta_{\cal H}(s-{\textstyle\frac{1}{2}}) - \zeta_{{\cal H}_0}(s-{\textstyle\frac{1}{2}}) \Big] \, \, .
\end{eqnarray*}
Following the same pattern as in Section \S.2 we regularize the $3$D vortex Casimir energy $\Delta E_V^C$ by using the spectral zeta function at a regular point in the $s$-complex plane
\begin{equation}
\Delta E_V^C({\cal K})(s) = \frac{\hbar \mu}{2} \Big( \frac{\mu^2}{m^2} \Big)^s \Big[ {\cal Z}_{\cal K}(s) - {\cal Z}_{{\cal K}_0}(s)  \Big] = \frac{\hbar \mu}{2} \Big(\frac{\mu^2}{m^2} \Big)^s \frac{m L}{2\sqrt{\pi}} \frac{\Gamma[s-\frac{1}{2}]}{\Gamma[s]} \Big[ \zeta_{\cal H} (s-\textstyle{\frac{1}{2}}) -\zeta_{{\cal H}_0} (s-\textstyle{\frac{1}{2}}) \Big] \label{rvsts}
\end{equation}
in such a way that in the limit $s\rightarrow - \frac{1}{2}$, which is a pole of $\Delta E_V^C({\cal K})(s)$, the physical response is recovered. Moreover, the contribution of the fermionic ghost particles, encoded in the spectrum of the PDO,
\[
{\cal H}^{G} = - \Delta +  |\psi^2| = - \frac{\partial^2}{\partial x_1^2} - \frac{\partial^2}{\partial x_2^2} - \frac{\partial^2}{\partial x_3^2} +  |\psi^2| \, \, ,
\]
must be subtracted, whereas fluctuations are accounted for, and must be added, by the spectrum of ${\cal L}_3$. Thus, the $3$D regularized vortex Casimir energy is the sum of these three contributions:
\[
\Delta E_V^C(s) = \Delta E_V^C({\cal K})(s) - \Delta E_V^C({\cal H}^G)(s) + \Delta E_V^C({\cal L}_3)(s)
\]
coming from the ${\cal K}$, ${\cal L}_3$ and ${\cal H}^G$-fluctuations. By regularizing also the contributions of the ghost and $a_3$ fluctuations by means of their spectral zeta functions
\begin{eqnarray*}
\Delta E_V^C({\cal K}^{\rm G})(s) &=& \frac{\hbar \mu}{2} \Big( \frac{\mu^2}{m^2} \Big)^s \Big[ {\cal Z}_{{\cal K}^G}(s) - {\cal Z}_{{\cal K}_0^G}(s) \Big] =
\frac{\hbar \mu}{2} \Big(\frac{\mu^2}{m^2} \Big)^s \frac{l}{2\sqrt{\pi}} \frac{\Gamma[s-\frac{1}{2}]}{\Gamma[s]} \Big[ \zeta_{{\cal H}_0^G} (s-\textstyle{\frac{1}{2}}) - \zeta_{{\cal H}^G} (s-\textstyle{\frac{1}{2}}) \Big] \\
\Delta E_V^C({\cal L}_3)(s) &=& \frac{\hbar \mu}{2} \Big( \frac{\mu^2}{m^2} \Big)^s \Big[ {\cal Z}_{{\cal L}_3}(s) - {\cal Z}_{{\cal L}_{30}}(s)\Big] = \frac{\hbar \mu}{2} \Big(\frac{\mu^2}{m^2} \Big)^s \frac{l}{2\sqrt{\pi}} \frac{\Gamma[s-\frac{1}{2}]}{\Gamma[s]} \Big[ \zeta_{{\cal L}_3} (s-\textstyle{\frac{1}{2}}) -\zeta_{{\cal L}_{30}} (s-\textstyle{\frac{1}{2}}) \Big]
\end{eqnarray*}
we obtain:
\begin{eqnarray}
\Delta E_V^C(s)&=&\frac{\hbar \mu}{2} \Big(\frac{\mu^2}{m^2} \Big)^s \frac{l}{2\sqrt{\pi}} \frac{\Gamma[s-\frac{1}{2}]}{\Gamma[s]} \Big[ \zeta_{\cal H} (s-\textstyle{\frac{1}{2}}) -\zeta_{{\cal H}_0} (s-\textstyle{\frac{1}{2}}) \Big] = \nonumber \\ &=&
\frac{\hbar \mu l}{4\sqrt{\pi}} \Big(\frac{\mu^2}{m^2} \Big)^s \cdot \frac{\Gamma[s-\frac{1}{2}]}{\Gamma[s]} \Big[  \frac{1}{4\pi \Gamma[s-\frac{1}{2}]} \sum_{n=1}^\infty \sum_{a=1}^4\frac{ [\mathbf{c}_n({\cal H}^+)]_{aa}}{u^{2s+2n-3}} \Gamma[s+n-{\textstyle\frac{3}{2}}]-2 N u^{-2s+1}\Big] \label{divregce}
\end{eqnarray}
because ${\cal H}^G$ and ${\cal L}_3$ are identical PDO's and thus the ghost and the third component vector potential fluctuations annihilate each other. We recall that $N$ is the vorticity of the vortex string.

Once we have derived (\ref{divregce}), a renormalization process must be implemented in order to tame the divergences of $\Delta E_V^C (s)$ at the physical limit $s\rightarrow -\frac{1}{2}$. Within the zeta function regularization procedure the more severe divergences appear in the lower three terms of the asymptotic expansion. To put into practice the renormalization procedure we distinguish between the contributions to
$\Delta E_V^C (s)$ of divergent and finite terms:
\[
\Delta E_V^C (s) = \Delta E_V^{C(1)}(s) +  \Delta E_V^{C(2)}(s) +  \Delta E_V^{C_3}(s) +  \Delta E_V^{C_{ZM}}(s) \, \, .
\]
Here
\begin{eqnarray*}
\Delta E_V^{C(1)}(s) &=& \frac{\hbar \mu l }{16\pi \sqrt{\pi}} \Big( \frac{\mu^2}{m^2} \Big)^s \cdot \frac{\Gamma[s-\frac{1}{2}]}{\Gamma[s]}  \sum_{a=1}^4 \frac{[\mathbf{c}_1({\cal H}^+)]_{aa}}{u^{2s-1}} \\
\Delta E_V^{C(2)}(s) &=& \frac{\hbar \mu l}{4\sqrt{\pi}} \Big( \frac{\mu^2}{m^2}\Big)^s \cdot \frac{\Gamma[s+\frac{1}{2}]}{4\pi \Gamma[s]} \sum_{a=1}^4 \frac{[\mathbf{c}_2({\cal H}^+)]_{aa}}{u^{2s+1}}
\end{eqnarray*}
refer respectively to the contribution of the first and second Seeley coefficients in the asymptotic series formula of the vortex Casimir energy $\Delta E_V^C(s)$. Of course, the contribution of the ${\rm tr}\,\textbf{c}_0({\cal H}^+)$ would be even more divergent, but it does not appear in the vortex Casimir energy because it is canceled by the contribution of the vacuum zeta function ${\cal Z}({\cal H}_0)(-1/2)$, i.e., by zero point renormalization. The interesting facts to be pointed out about the divergences of the $3$D vortex string Casimir energy are: (1) $\Delta E_V^{C(1)}(-1/2)$
has a divergence proportional to $\Gamma(-1)$. (2) The divergence of $\Delta E_V^{C(2)}(-1/2)$ arises as the pole of $\Gamma(s)$ at $s=0$. (3) Factors respectively of $u^2$ and $u^0$ in these lower two terms of the series tell us that these contributions would survive in the infinite mass limit. Therefore, the divergences coming from massive fluctuations, i.e., appearing in factors of the old Seeley coefficients, must be exactly cancelled according to the heat kernel renormalization criterion. Moreover, the exponents of $u$ encode in the spectral zeta function the standard divergences of QFT: for instance, divergences coming from the $c_1$ coefficients correspond to quadratic divergences in the Feynman graphs when a momentum cutoff is used, those appearing in $c_2$ contributions come from QFT logarithmic divergences{\footnote{The stronger divergences, quartic in $3$D, are associated with vacuum energies, i.e., with $c_0$ coefficients that are proportional to $u^4$. Note also that in the zeta function regularization procedure these quartic divergences reappear in the disguise of $\Gamma(-2)$. Fortunately, these quartic divergences are supressed by zero point renormalization.}}.

The remaining summands in the series, however,
\[
\Delta E_V^{C_3}(s) = \frac{\hbar \mu l }{16\pi \sqrt{\pi}} \Big( \frac{\mu^2}{m^2} \Big)^s \cdot\frac{1}{\Gamma[s]}  \sum_{n=3}^\infty \sum_{a=1}^4 \frac{[c_n({\cal H})]_{aa}}{u^{2s+2n-3}} \Gamma[s+n- {\textstyle\frac{3}{2}} ]
\]
are finite at $s=-1/2$ and proportional to negative powers of $u$, a fact that tells us that they escape from the need of renormalization. The zero mode contribution, however, survives even in the infinite mass limit
but it is divergent at the physical value of the $s$ complex parameter. Indeed,
\[
\Delta E_V^{C_{ZM}}(s) = -\frac{\hbar \mu  l}{2\sqrt{\pi}} \Big( \frac{\mu^2}{m^2} \Big)^s \cdot \frac{\Gamma[s-\frac{1}{2}]}{\Gamma[s]} \, N \, u^{-2s+1}
\]
is divergent at $s=-1/2$ because $\Gamma(s-1/2)$ has a pole there. It is of note that this contribution is proportional to twice the vorticity $2N$, a number that counts the zero modes.

In order to fix the renormalizations needed it is convenient a closer analysis of the vortex Casimir energy divergences near the dangerous pole at $s=-1/2$. A power expansion of the divergent contributions in the neighborhood of this point shows the just mentioned structure:
\begin{eqnarray*}
\Delta E_V^{C(1)}(s) &=& \hbar \mu \Big( \frac{\mu^2}{m^2} \Big)^s \frac{m L}{16 \pi \sqrt{\pi}} \Big( \frac{1}{2\sqrt{\pi}(s+\frac{1}{2})} + \frac{1-\gamma-\psi(-\frac{1}{2})}{2\sqrt{\pi}} + o(s+{\textstyle\frac{1}{2}}) \Big) \sum_{a=1}^4 [c_1({\cal H}^+)]_{aa} u^2 \\
\Delta E_V^{C(2)}(s) &=& \hbar \mu \Big( \frac{\mu^2}{m^2} \Big)^s \frac{m L}{16 \pi \sqrt{\pi}} \Big( \frac{-1}{2\sqrt{\pi}(s+\frac{1}{2})} + \frac{\gamma+\psi(-\frac{1}{2})}{2\sqrt{\pi}} + o(s+{\textstyle\frac{1}{2}}) \Big) \Big( \Big\langle \Sigma_2(\psi,V_\alpha)\Big\rangle  -8\pi N u^2 \Big) \\
\Delta E_V^{C_{ZM}}(s) &=& -\hbar \mu \Big( \frac{\mu^2}{m^2} \Big)^s \frac{m L}{2 \sqrt{\pi}} \Big( \frac{1}{2\sqrt{\pi}(s+\frac{1}{2})} + \frac{1-\gamma-\psi(-\frac{1}{2})}{2\sqrt{\pi}} + o(s+{\textstyle\frac{1}{2}}) \Big) \,N u^2
\end{eqnarray*}
where $\gamma$ is the Euler Gamma constant and $\psi(s)$ is the Digamma function. The second Seeley coefficient has been split into two summands
\[
\sum_{a=1}^4 [c_2({\cal H^+})]_{aa} =\Big\langle\Sigma_2(\psi,V_\alpha)\Big\rangle-8\pi N u^2
\]
distinguishing between the zero mode contribution $-8\pi N$ and the contribution of the vortex fields expressed in terms of the old second Seeley coefficient, that is, derived in the standard GdW procedure, proportional to $\Big\langle \Sigma_2 (\psi, V_\alpha)\Big\rangle$ where $\Sigma_2$ is:
{\small\[
\Sigma_2(\psi,V_\alpha) = -\frac{5}{6} \Delta |\psi|^2 - \frac{1}{3} \Delta (\sum_{\alpha=1}^2V_\alpha V_\alpha) + 4 \sum_{\alpha,\beta=1}^2 (D_\alpha \psi_\beta)^2 + \frac{13}{4} (1-|\psi|^2)^2 - 2\sum_{\alpha=1}^2V_\alpha V_\alpha(1-|\psi|^2) + \frac{1}{3} (\sum_{\alpha=1}^2V_\alpha V_\alpha)^2 \, \, .
\]}
All the singular contributions to the vortex Casimir energy can be rearranged in the form:
\begin{eqnarray*}
&& \Delta E_V^{C(1)}(s)+\Delta E_V^{C(2)}(s)+ \Delta E_V^{C_{ZM}}(s)\simeq_{s\to -1/2} \\
&& \simeq_{s\to -1/2} \left\{ \hbar \mu \Big( \frac{\mu^2}{m^2} \Big)^s \frac{m L}{16 \pi \sqrt{\pi}} \Big( \frac{1}{2\sqrt{\pi}(s+\frac{1}{2})} + \frac{1-\gamma-\psi(-\frac{1}{2})}{2\sqrt{\pi}} + o(s+{\textstyle\frac{1}{2}}) \Big) \sum_{a=1}^4 [c_1({\cal H}^+)]_{aa} u^2 + \right. \\  && + \hbar \mu \Big( \frac{\mu^2}{m^2} \Big)^s \frac{m L}{16 \pi \sqrt{\pi}} \Big( \frac{-1}{2\sqrt{\pi}(s+\frac{1}{2})} + \frac{\gamma+\psi(-\frac{1}{2})}{2\sqrt{\pi}} + o(s+{\textstyle\frac{1}{2}}) \Big) \Big\langle \Sigma_2(\psi,V_\alpha)\Big\rangle- \\ && \left. -\hbar \mu \Big( \frac{\mu^2}{m^2} \Big)^s \frac{m L}{2 \sqrt{\pi}} \Big( \frac{1}{2\sqrt{\pi}(s+\frac{1}{2})} + \frac{1-\gamma-\psi(-\frac{1}{2})}{2\sqrt{\pi}}- \frac{1}{2\sqrt{\pi}(s+\frac{1}{2})}+ \frac{\gamma+\psi(-\frac{1}{2})}{2\sqrt{\pi}}+ o(s+{\textstyle\frac{1}{2}}) \Big) \,N u^2 \right\}
\end{eqnarray*}
There appear three types of singularities that need to be cancelled: (1) In the first line the divergences appear in the contribution to the vortex Casimir energy of the first Seeley coefficients. The heat kernel renormalization criterion
demands exact cancellation of this divergent term proportional to $u^2$ by subtracting the appropriate contribution to the energy of some mass renormalization counter-terms. In particular a minimal renormalization scheme must be implemented to tame the quadratic
divergences of the Higgs tadpole plus the self-energy graphs of all the scalar and vector bosons, as well as the fermionic ghosts. Use of the vacuum spectral zeta function is convenient to regularize the pertinent divergent graphs. We will not develop this delicate procedure
here, see \cite{Alonso2004:prd,Alonso2005:prd,Alonso2006:hepth, Mateos2009:pos, Alonso2008:npb} to see how this renormalization works in the superenormalizable, henceforth, easier planar AHM. Simply we shall take equal to zero
the contribution written in the first line legitimated by the heat kernel renormalization criterion {\footnote{We remark that ${\rm tr} \,\textbf{c}_1({\cal H}^+)$, like  ${\rm tr} \, \textbf{c}_0({\cal H}^+)$, is infrared divergent, although only as $\log L^2$. Mass renormalization takes care also of this infrared divergence.}}. (2) The same situation happens with the divergent contributions in the second line coming from the old second Seeley coefficient because it is proportional to $u^0$ and survives in the infinite mass limit. The divergences, even being smoother, are more involved. One must cope now with the subdominant logarithmic divergences of the graphs just mentioned plus the logarithmic divergences of one-loop graphs with three Higgs legs plus all the one-loop graphs with four external legs of the fields working in the AHM. This means that we shall use the energies
due to the counter-terms arising in the coupling constant and wave function {\footnote{The terms which are field derivatives in $\Sigma_2$ are exactly canceled by wave function renormalization of the scalar and vector massive particles.}} renormalizations adjusted to exactly cancel the contribution in the second line. (3) In the third line we observe an exact cancelation between the divergences due to the zero modes. There is, however, a finite remnant that must be kept because the heat kernel renormalization criterion does not apply to massless fluctuations.

Finally the one-loop vortex mass shift per length unit is obtained by taking the limit $s\rightarrow -\frac{1}{2}$ in the sum of the finite remnant of the whole zero mode contribution plus the partial sum up to $n_T$ order
in the series of finite terms $\Delta E_V^{C_3}(s)$ taking of course the physical value $u=1$:
\begin{equation}
\frac{\Delta E_V^{C}}{L} = - \frac{\hbar m^2}{32\pi^2}   \sum_{n=3}^{n_T} \sum_{a=1}^4 [c_n({\cal H}^+)]_{aa} \Gamma[n- 2 ] -\frac{\hbar m^2}{4\pi }  \,N \label{quantumshift3} \, \, .
\end{equation}
This energy per unit length is precisely the one-loop string tension shift induced in the BPS vortices by quantum fluctuations.

In Table 4 we display the responses obtained from this formula up to $n_T=6$ for several values of the vorticity $N$. The last row offers the best estimations of the $N$-vortex string tension quantum corrections. The necessary Seeley coefficients were previously displayed in Table 1.

\begin{table}[h]
\centering\begin{tabular}{|c|c|c|c|c|c|} \hline
$n_T$ & $\rule[-0.3cm]{0cm}{0.9cm} \frac{\Delta E_V^{N=1}}{\hbar m^2 L}$ & $\frac{\Delta E_V^{N=2}}{\hbar m^2 L}$ & $\frac{\Delta E_V^{N=3}}{\hbar m^2 L}$ & $\frac{\Delta E_V^{N=4}}{\hbar m^2 L}$ & $\frac{\Delta E_V^{N=5}}{\hbar m^2 L}$ \\ \hline\hline
$3$ & $-0.0815$ & $-0.1612$ & $-0.2342$ & $-0.2995$ & $-0.3566$ \\
$4$ & $-0.0818$ & $-0.1604$ & $-0.2297$ & $-0.2882$ & $-0.3354$ \\
$5$ & $-0.0820$ & $-0.1597$ & $-0.2265$ & $-0.2806$ & $-0.3214$ \\
$6$ & $-0.0821$ & $-0.1591$ & $-0.2240$ & $-0.2749$ & $-0.3112$ \\ \hline
\end{tabular}\hspace{1cm}\begin{tabular}{c} \includegraphics[height=3cm]{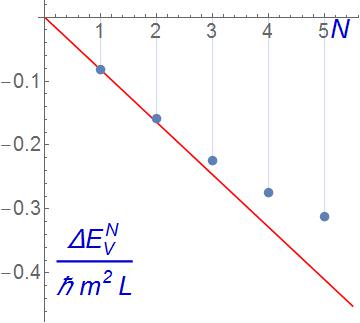}\end{tabular}
\caption{\it Estimation of the quantum correction to the $N$-vortex filament string tension up to vorticity $N=5$ computed from the $3\leq n_T\leq 3$ partial sums of the series (\ref{quantumshift3}).}
\end{table}

\section{Conclusions and further comments}

From the results in this work we draw two main conclusions:

\begin{itemize}

\item The modified Gilkey-de Witt heat kernel expansion designed in References \cite{Alonso2012:epjc} and \cite{Alonso2014:jhep} to control the impact
of zero modes in the calculations of quantum corrections to kink masses and domain wall surface tensions due to one-loop fluctuations
in scalar field theory has been successfully generalized to analyze one-loop fluctuations of both planar and cylindrical BPS vortices in the Abelian Higgs model.

\item The new estimations are more precise than those obtained in \cite{Alonso2004:prd} and \cite{Alonso2005:prd} by using the standard Gilkey-deWitt expansion. The archive of new data clearly suggests that weak repulsive forces between BPS vortices arise caused by the one-loop vortex fluctuations. In extended $N=2$ supersymmetry, however, the one-loop vortex mass shift and the central charge are adjusted in such a way that one may say the BPS bound is preserved at the quantum level, see \cite{Rebhan2004:npb}. Thus, one may conclude that some degree of extended supersymmetry is needed in order to preserve the BPS character of topological solitons in the quantum domain.

\end{itemize}
We stress that our calculations have been performed over a dilute gas of vortices with a few number of quanta of magnetic flux spread over the whole plane. In Reference \cite{Ferreiros2014:prd}, however, a different arrangement of vortices has been analyzed. The authors addressed the quantization of a bunch of magnetic flux quanta in a parallelogram, a normalization square, such that the Bradlow limit was almost reached. This means that, after imposing quasi-periodic boundary conditions on the fluctuations, the magnetic flux of the vortex configuration is very close to the area of the equivalent genus one Riemann surface. Exactly at the Bradlow limit the zero modes form the first Landau level of the Landau problem posed in this Riemann surface and a reshaping of the work of Ferreiros at al from the point of view proposed in this paper will be probably rewarding. Although the new technique has been designed to deal with one-loop fluctuations or vacuum energies of low dimensional topological solitons one may speculate with its application to other extended objects supporting zero modes of fluctuation. For instance, it is tempting to try this quantization method on the BPS magnetic monopoles of the bosonic sector in the ${\cal N}=2$ SUSY gauge theory of Seiberg and Witten, see \cite{Seiberg1994:npb}, and compare the results obtained with those achieved  in the supersymmetric framework in \cite{Rebhan2006:jhep}.

We have successfully applied the improved zeta function procedure in calculations of domain wall surface tension \cite{Alonso2014:jhep} and in the regularization of tunnel determinants in quantum mechanics, see \cite{Alonso2014:aip}. It seems plausible that the new method may be also effective in the analysis of tunnel determinants appearing in connection with Yang-Mills and/or gravitational instantons, see \cite{Hawking1977:cmp,Belavin1975:plb,Hooft1986:pr,Gibbons1978:plb,Eguchi1978:plb}. Other objects of the greatest physical interest as black holes may be understood as solitons, see e.g. \cite{Salam1976:plb}. Thus, our method is of potential interest in dealing with quantum fields in the background of solitonic black holes. To finish, one might think about the applicability of the improved Gilkey-deWitt expansion to more exotic topological solitons as, for instance, the BPS vortices of two species arising in the gauged non-linear $\mathbb{CP}^N$ \cite{Alonso2015:jhep}, or, to compactons appearing in models with higher-order kinetic terms, see \cite{Bazeia2015:prd} where one-loop correction to their classical masses have been computed.

\section*{Acknowledgements}

The authors acknowledge the Spanish Ministerio de Econom\'{\i}a y Competitividad for financial support under grant MTM2014-57129-C2-1-P. They are also grateful to the Junta de Castilla y Le\'on
for financial help under grant VA057U16.

%\bibliography{PaperBiblio}
%\bibliographystyle{unsrt}

\end{document}